\documentclass[12pt]{article}

\font\grande=cmr9.5 scaled \magstep4
\font\medio=cmr9.5 scaled \magstep2
\outer\def\beginsection#1\par{\medbreak\bigskip
      \message{#1}\leftline{\bf#1}\nobreak\medskip
\vskip-\parskip
      \noindent}
\usepackage{graphicx} 
\begin{document}
\bibliographystyle {unsrt}

\titlepage

\begin{flushright}
CERN-PH-TH/2011-287
\end{flushright}

\vspace{10mm}
\begin{center}
{\grande Compressible hydromagnetic nonlinearities}\\
\vspace{0.5cm}
{\grande in the predecoupling plasma}\\
\vspace{1.5cm}
 Massimo Giovannini\footnote{Electronic address: massimo.giovannini@cern.ch}\\
\vspace{1cm}
{{\sl Department of Physics, 
Theory Division, CERN, 1211 Geneva 23, Switzerland }}\\
\vspace{0.5cm}
{{\sl INFN, Section of Milan-Bicocca, 20126 Milan, Italy}}
\vspace*{0.5cm}
\end{center}

\vskip 0.5cm
\centerline{\medio  Abstract}
\vskip 0.5cm
The adiabatic inhomogeneities of the scalar curvature lead to a compressible flow affecting the dynamics of the hydromagnetic nonlinearities. The influence of the plasma on the evolution of a putative magnetic field is explored with the aim of obtaining an effective description 
valid for sufficiently large scales. The bulk velocity of the plasma, computed in the framework of the $\Lambda$CDM scenario, feeds back into the evolution of the magnetic power spectra leading to a (nonlocal) master equation valid in Fourier space and similar to the ones discussed in the context of wave turbulence. Conversely, in physical space, the magnetic power spectra obey a  Schr\"odinger-like equation whose effective potential depends on the large-scale curvature perturbations. Explicit solutions are presented both in physical space and in Fourier space. It is argued that curvature inhomogeneities, compatible with the WMAP 7yr data, shift to lower wavenumbers the magnetic diffusivity scale. 
\noindent

\vspace{5mm}

\vfill
\newpage
\renewcommand{\theequation}{1.\arabic{equation}}
\setcounter{equation}{0}
\section{Introduction}
\label{sec1}
The evolution equations of the magnetic power spectra in a conducting fluid have been the subject of extended studies with particular attention to the regime of high Reynolds numbers (see, e.g. \cite{moffat,parker,zeldovich,biskamp}).  In standard treatments the flow is often passive (in the sense that it is given as an external input of the problem), it is isotropic and and it is assumed to be incompressible in various situations ranging from the usual one-fluid description provided by magnetohydrodynamics (see, e.g. \cite{biskamp}) to the applications of gyrotropic turbulence \cite{zeldovich}.
Exceptions to the previous statement are, for instance, the compressible turbulence \cite{yaglom} or the hydromagnetic evolution in the presence of acoustic disturbances.

Prior to photon decoupling, the physical properties of the primeval plasma can be directly scrutinized 
by means of the measurements of the Cosmic Microwave Background (CMB) temperature and polarization anisotropies. These observations 
show, with an accuracy which is sufficient for the present considerations, the predominance of the adiabatic curvature perturbations over any other (possibly subleading) entropic contribution \cite{NAD1,NAD2} (see also \cite{rev} for a recent review). Indeed, the results of the WMAP observations (see, e.g. 
the WMAP 7yr data release \cite{wmap7a,wmap7b,wmap7c})  imply that the position of the first acoustic peak of the temperature autocorrelations and the position of the first anticorrelation peak in the temperature-polarization 
power spectra are in a fixed ratio (i.e. approximately $3/4$). Such a ratio can be easily derived  
by assuming a dominant adiabatic component of curvature inhomogeneities in the initial conditions of 
the Einstein-Boltzmann hierarchy.  Tensor and vector 
modes play a negligible role (see \cite{gw1,gw2} and references therein) but are anyway absent in the vanilla $\Lambda$CDM scenario (where $\Lambda$ stands for the dark energy component and CDM for the cold dark matter component).  The adiabatic nature of the large-scale curvature perturbations determines also the statistical properties of the fluid flow: the baryon-photon and the cold dark matter velocities are Gaussian, isotropic and irrotational. The bulk velocity of the plasma (i.e.  $\vec{v}_{\mathrm{b}}$  in what follows) coincides, at early times,  with the baryon velocity \cite{max1} (see also \cite{c1}). When photons and baryons are tightly coupled the photon-baryon flow is also compressible and the typical amplitude of the corresponding velocity field is determined by the large-scale curvature perturbations. Finally the correlation functions of the bulk velocity of the plasma is exponentially suppressed for large times beyond a typical time scale, denoted by $\tau_{\mathrm{d}}$ and related to the diffusive (or Silk) damping. 

The main motivation of this paper can be be summarized as follows.
Prior to decoupling the Prandtl number (i.e. $Pr_{\mathrm{magn}} = R_{\mathrm{magn}}/R_{\mathrm{kin}}$) is 
$Pr_{\mathrm{magn}} = {\mathcal O}(10^{20})$ with $R_{\mathrm{kin}} < 1$ and $R_{\mathrm{magn}} \gg 1$. In such a system the fluid flow is  computable (for instance in one of the popular versions 
of the $\Lambda$CDM scenario) and, in some sense, even accessible 
experimentally. Consequently, a natural question to ask is wether it is possible 
to account for the effect of the large-scale flow on the evolution of the magnetic power spectra. This problem is formally analog to (but physically very different from) some of the themes mentioned in the first paragraph of this 
introductory section. In the recent past close attention has been paid to the analysis of the effects of large-scale magnetic fields on the evolution of curvature perturbations and on the CMB anisotropies (see, e.g., \cite{max1,max2} and references therein). In this analysis the complementary (and to some extent inverse) point of view will be taken:  given the properties of the large-scale 
flow in the framework of the $\Lambda$CDM paradigm we ought to know how is the evolution of a putative magnetic field modified, in particular for large-scales.

A similar problem 
arises in the statistical treatment of conducting fluids \cite{moffat,parker, zeldovich}
where an interesting approach has been pioneered by Kazantsev \cite{kaza}, Kraichnan and Nagarajan \cite{kraichnan3} (see also \cite{kraichnan1,naga}) and Vainshtein \cite{vain0}. The common aspect of the approaches of 
\cite{kaza,kraichnan3,vain0} is that the flow is assumed Gaussian, isotropic and incompressible: these are also
the three main physical analogies with the situation addressed in this paper. 

In spite of these relevant analogies there also two important differences. The first one is that 
the fluctuations of the geometry (as well as the effects related to the expansion of the background) do not play any role 
in the hydromagnetic treatment of conducting fluids; conversely, prior to decoupling, the large-scale flow is exactly determined by the fluctuations of the spatial curvature. In this sense the flow discussed in the present analysis is not passive but rather computed from the large-scale curvature perturbations. The second physical difference concerns the hierarchy between the Reynolds numbers which are 
simultaneously high (i.e. $R_{\mathrm{kin}} \gg 1$ and $R_{\mathrm{magn}} \gg 1$) in the standard 
hydromagnetic treatments\footnote{ In spite of the fact that the three approaches of \cite{kaza, kraichnan3,vain0} 
are conceptually related, the analysis of Kazantsev \cite{kaza} seems more appropriate for the present ends and 
will therefore constitute the basis for the generalizations developed in the present paper.} \cite{kaza} (see also   \cite{kraichnan3,vain0}). This is not the case after electron-positron annihilation and the smallness of $R_{\mathrm{kin}}$ is partly related to the properties of the bulk flow whose evolution is described within the same perturbative treatment used for the large-scale curvature fluctuations.

The layout of the present paper is the following. In section \ref{sec2} the set of governing equations will be introduced in its general form. In section \ref{sec3} the 
$\Lambda$CDM fluid flow will be fed back into the evolution of the magnetic fields leading to an explicit integrodifferential equation valid  for a compressible flow. The evolution of the magnetic power spectra is derived both in Fourier space and in physical space (section \ref{sec4}). Explicit solutions are derived in section 
\ref{sec5}.  Section \ref{sec6} contains the concluding remarks. In the appendixes useful technical results have been collected for the interested readers.

\renewcommand{\theequation}{2.\arabic{equation}}
\setcounter{equation}{0}
\section{Governing equations}
\label{sec2}
The hydromagnetic equations in the presence of the fluctuations of the geometry have been deduced in different gauges (see, e.g.  \cite{max1,max2} and references therein). Here the essentials of the curved space 
description of the predecoupling plasma will be briefly reviewed. The approach of this paper is, in some sense, inverse of the analysis of 
\cite{max1,max2}:  we are here concerned with the problem of deducing an averaged description for the evolution of the magnetic power spectra when the flow is determined in the framework of the $\Lambda$CDM paradigm. 
\subsection{Curvature perturbations}
In a conformally flat background metric $\overline{g}_{\mu\nu}= a^2(\tau)\eta_{\mu\nu}$ the relativistic fluctuations of the 
geometry are given by
\begin{equation}
\delta_{\mathrm{s}} g_{00}(\vec{x},\tau) = 2\, a^2(\tau)  \phi(\vec{x},\tau),\qquad \delta_{\mathrm{s}} g_{ij}(\vec{x},\tau) = 2\,a^2(\tau) 
\psi(\vec{x},\tau) \delta_{ij},
\label{met1}
\end{equation}
in the longitudinal coordinate system. In Eq. (\ref{met1}) $\delta_{\mathrm{s}}$ denotes a metric perturbation which preserves the scalar nature of the 
fluctuation since, in the $\Lambda$CDM paradigm, the dominant source of inhomogeneity 
comes from the scalar modes of the geometry while the tensor and the vector inhomogeneities are absent. The scalar inhomogeneities are customarily parametrized either in terms of the curvature 
flutuations on comoving orthogonal hypersurfaces (conventionally denoted with ${\mathcal R}$) 
or in terms of the curvature perturbations on uniform density hypersurfaces (conventionally denoted by $\zeta$).
Both ${\mathcal R}$ and $\zeta$ can be defined in terms of the variables $\phi$ and $\psi$ appearing 
in Eq. (\ref{met1}):
\begin{equation}
{\mathcal R} = - \psi - \frac{{\mathcal H} ({\mathcal H} \phi + \partial_{\tau} \psi)}{{\mathcal H}^2 - \partial_{\tau}{\mathcal H}}, \qquad \zeta = {\mathcal R} + \frac{\nabla^2 \psi}{3 ({\mathcal H}^2 - \partial_{\tau}{\mathcal H})},
\label{met2}
\end{equation}
where ${\mathcal H} = a H$ and $ H$ is the usual Hubble rate. The variables ${\mathcal R}$ and $\zeta$ are gauge-invariant and the second relation 
of Eq. (\ref{met2}) stems from the Hamiltonian constraint written in 
the longitudinal gauge of Eq. (\ref{met1}). 
For a more thorough definition of large-scale curvature perturbations in different coordinate systems see \cite{c1,c2,c3} and references therein.
The large-scale curvature perturbations are then assigned in terms of the two-point function of ${\mathcal R}_{*}$, in Fourier space and prior to matter-radiation equality where curvature perturbations are approximately constant, at 
least  in the case of the $\Lambda$CDM paradigm: 
\begin{equation}
\langle {\mathcal R}_{*}(\vec{k}) {\mathcal R}_{*}(\vec{p}) \rangle = \frac{2\pi^2}{k^3} {\mathcal P}_{{\mathcal R}}(k) 
\delta^{(3)}(\vec{k} + \vec{p}), \qquad 
{\mathcal P}_{\mathcal R}(k) = {\mathcal A}_{{\mathcal R}} \biggl(\frac{k}{k_{\mathrm{p}}}\biggr)^{n_{\mathrm{s}} -1}.
\label{met3}
\end{equation}
In Eq. (\ref{met3}) ${\mathcal A}_{{\mathcal R}} = 2.43 \times 10^{-9}$ is the amplitude of the power spectrum 
at the pivot scale $k_{\mathrm{p}} =0.002\, \mathrm{Mpc}^{-1}$ and $n_{s} = 0.963$ is, by definition, the scalar spectral index. The numerical 
values of ${\mathcal A}_{{\mathcal R}}$ and $n_{\mathrm{s}}$ are determined from the WMAP 7yr data 
analyzed in the light of the vanilla $\Lambda$CDM scenario; the values of the remaining parameters are given by \cite{wmap7a,wmap7b,wmap7c}:
\begin{equation}
( \Omega_{\mathrm{b}0}, \, \Omega_{\mathrm{c}0}, \Omega_{\mathrm{de}0},\, h_{0},\,n_{\mathrm{s}},\, \epsilon_{\mathrm{re}}) \equiv (0.0449,\, 0.222,\, 0.734,\,0.710,\, 0.963,\,0.088),
\label{R3a}
\end{equation}
where, as usual, $\Omega_{x \,0}$ denotes the present critical fraction of the species $x$; $h_{0}$ is the  Hubble constant in units of $100\, \mathrm{km}/\mathrm{sec}/\mathrm{Mpc}$ and $\epsilon_{\mathrm{re}}$ is the optical depth 
to reionization. 
\subsection{Two-fluid equations}
The comoving electromagnetic fields and the comoving concentrations of electrons and ions are given by
\begin{eqnarray}
&& \vec{E}(\vec{x},\tau) = a^2(\tau) \vec{{\mathcal E}}(\vec{x},\tau), \qquad  \vec{B}(\vec{x},\tau) = a^2(\tau) \vec{{\mathcal B}}(\vec{x},\tau),
\nonumber\\
&& n_{\mathrm{i}}(\vec{x},\tau) = a^3(\tau) \tilde{n}_{\mathrm{i}}(\vec{x},\tau),\qquad 
n_{\mathrm{e}}(\vec{x},\tau) = a^3(\tau) \tilde{n}_{\mathrm{e}}(\vec{x},\tau).
\label{S1}
\end{eqnarray}
Thus Maxwell's equations read
\begin{eqnarray}
&& \vec{\nabla} \cdot \vec{E} = 4 \pi e (n_{\mathrm{i}} - n_{\mathrm{e}}),\qquad\vec{\nabla} \cdot \vec{B} =0,
\label{S2}\\
&& \vec{\nabla} \times \vec{E} = - \partial_{\tau} \vec{B}, \qquad  \vec{\nabla}\times \vec{B} = 4\pi e (n_{\mathrm{i}}\, \vec{v}_{\mathrm{i}} - 
n_{\mathrm{e}}\, \vec{v}_{\mathrm{e}} ) + \partial_{\tau}\vec{E}.
\label{S3}
\end{eqnarray}
The velocities of the electrons, ions and photons obey, respectively, the following 
set of coupled equations:
\begin{eqnarray}
&& \partial_{\tau}\vec{v}_{\mathrm{e}} + {\mathcal H}\,\vec{v}_{\mathrm{e}} = - \frac{e n_{\mathrm{e}}}{\rho_{\mathrm{e}} \, a^{4}} [ \vec{E} + \vec{v}_{\mathrm{e}} \times \vec{B}] - \vec{\nabla} \phi 
+ 
\frac{4}{3} \frac{\rho_{\gamma}}{\rho_{\mathrm{e}}} a 
\Gamma_{\gamma \, \mathrm{e}} (\vec{v}_{\gamma} - \vec{v}_{\mathrm{e}}) + a \Gamma_{\mathrm{e\,i}} ( \vec{v}_{\mathrm{i}} - \vec{v}_{\mathrm{e}}),
\label{SA}\\
&&  \partial_{\tau} \vec{v}_{\mathrm{i}} + {\mathcal H}\,\vec{v}_{\mathrm{i}} =   \frac{e n_{\mathrm{i}}}{\rho_{\mathrm{i}} \, a^{4}}[ \vec{E} + \vec{v}_{\mathrm{i}} \times \vec{B}] - \vec{\nabla} \phi 
+ 
\frac{4}{3} \frac{\rho_{\gamma}}{\rho_{\mathrm{i}}} a 
\Gamma_{\gamma \, \mathrm{i}} (\vec{v}_{\gamma}-\vec{v}_{\mathrm{i}} ) + a \Gamma_{\mathrm{e\,i}} \frac{\rho_{\mathrm{e}}}{\rho_{\mathrm{i}}}( \vec{v}_{\mathrm{e}} - \vec{v}_{\mathrm{i}}),
\label{SB}\\
&& \partial_{\tau} \vec{v}_{\gamma} = - \frac{1}{4} \vec{\nabla} \delta_{\gamma} - \vec{\nabla} \phi 
+ a \Gamma_{\gamma\mathrm{i}} (\vec{v}_{\mathrm{i}} - \vec{v}_{\gamma}) + 
a \Gamma_{\gamma\mathrm{e}}  ( \vec{v}_{\mathrm{e}} - \vec{v}_{\gamma}).
\label{SC}
\end{eqnarray}
In Eqs. (\ref{SA})--(\ref{SC}) the relativistic fluctuations of the geometry are included from the very beginning in terms of the longitudinal gauge variables of Eq. (\ref{met1}); the electron-photon,  electron-ion and ion-photon 
rates of momentum exchange appearing in Eqs. (\ref{SA})--(\ref{SC}) are given by\footnote{Note that $\Lambda_{\mathrm{C}}$ is the Coulomb logarithm \cite{spitzer,krall}.}:
\begin{eqnarray}
&&\Gamma_{\gamma\mathrm{e}} = \tilde{n}_{\mathrm{e}} 
\sigma_{\mathrm{e}\gamma},\qquad 
\Gamma_{\gamma\mathrm{i}} = \tilde{n}_{\mathrm{i}} 
\sigma_{\mathrm{i}\gamma},\qquad \sigma_{\mathrm{e}\gamma} 
= \frac{8}{3}\pi \biggl(\frac{e^2}{m_{\mathrm{e}}}\biggr)^2, \qquad 
\sigma_{\mathrm{i}\gamma} 
= \frac{8}{3}\pi \biggl(\frac{e^2}{m_{\mathrm{i}}}\biggr)^2,
\label{S9}\\
&& \Gamma_{\mathrm{e\,i}} = \tilde{n}_{\mathrm{e}} \sqrt{\frac{T}{m_{\mathrm{e}}}} \, \sigma_{\mathrm{e\,i}} = \Gamma_{\mathrm{i\, e}},\qquad \sigma_{\mathrm{e\,i}} = 
\frac{e^4}{T^2} \ln{\Lambda_{\mathrm{C}}},\qquad \Lambda_{\mathrm{C}} = \frac{3}{2 e^3} \sqrt{\frac{T^3}{\tilde{n}_{\mathrm{e}}\pi}}.
\label{S10}
\end{eqnarray}
In Eq. (\ref{S9}) and (\ref{S10}), $T$ and $\tilde{n}$ are, respectively,  
physical temperatures and physical concentrations\footnote{If the rates and the cross sections are expressed in terms of comoving temperatures $\overline{T} = a T$ and 
comoving concentrations $n = a^3 \, \tilde{n}$ the corresponding rates will inherit a scale factor for each 
mass. For instance $a\Gamma_{\mathrm{e\,i}}$ becomes $n_{\mathrm{e}} \, \sqrt{\overline{T}/(m_{\mathrm{e}} a)} \, (e^4/\overline{T}^2) \, \ln{\Lambda_{\mathrm{C}}}$,  if comoving temperature and concentrations are used. }.

\subsection{The problem of the closure}
The two global variables entering the one-fluid description are the center of mass velocity of the electron 
ion system and the total current, i.e. 
\begin{equation}
\vec{v}_{\mathrm{b}} = \frac{m_{\mathrm{e}}\vec{v}_{\mathrm{e}} 
+ m_{\mathrm{i}} \vec{v}_{\mathrm{i}}}{m_{\mathrm{e}} + m_{\mathrm{i}}}, \qquad 
\vec{J} = n_{\mathrm{i}} \vec{v}_{\mathrm{i}} - n_{\mathrm{e}} \vec{v}_{\mathrm{e}}.
\label{def1}
\end{equation}
The one-fluid equations supplemented by the incompressible closure are often 
used as a starting point for the analysis especially in connection with 
the potential emergence of interesting scaling laws
(see, e.g. \cite{olesen1}). The bulk velocity of the plasma coincides, in the latter case, with $\vec{v}_{\mathrm{b}}$ and the incompressibility condition dictates $\vec{\nabla}\cdot \vec{v}_{\mathrm{b}} =0$. Having said this, the reduction from the two-fluid to the one-fluid variables 
is essential, in the present context, exactly because $\vec{\nabla}\cdot\vec{v}_{\mathrm{b}} \neq 0$ and the velocity 
correlators depend on the density fluctuations of the plasma. 
The difference of Eqs. (\ref{SA}) and (\ref{SB}) leads to the evolution equation of the total current. Since 
the rate of Coulomb scattering is much larger than the 
conformal time derivatives of the current, the Ohm equation can be approximated by the standard form of the Ohm law (see fifth article of Ref. \cite{max1}):
\begin{equation}
\vec{J} = \sigma\biggl( \vec{E} + \vec{v}_{\mathrm{b}} \times \vec{B} + \frac{\vec{\nabla} p_{\mathrm{e}}}{e n_{0}} - 
\frac{\vec{J} \times \vec{B}}{e n_{0}} \biggr),
\label{pd5a}
\end{equation}
where $n_{0} = a^3 \tilde{n}_{\mathrm{e\,i}}$ denotes the common 
value of the (common) electron-ion concentration; $\sigma$ is the conductivity 
of the predecoupling plasma. After electron-positron annihilation 
$\sigma(\overline{T})$ can be estimated as 
\begin{equation}
\sigma(\overline{T}) = \sigma_{1}\frac{\overline{T}}{\alpha_{\mathrm{em}}}  \sqrt{\frac{\overline{T}}{m_{\mathrm{e}} a}} 
\frac{1}{\ln{\Lambda_{\mathrm{C}}}}, \qquad
 \Lambda_{\mathrm{C}}(\overline{T}) = \frac{3}{2 e^3} \biggl(\frac{\overline{T}^3}{\pi n_{\mathrm{e}}}\biggr)^{1/2} =  1.105\times 10^{8} \biggl(\frac{\omega_{\mathrm{b}0}}{0.02258}\biggr)^{-1/2},
 \label{R5}
\end{equation}
where $\sigma_{1} = 9/(8 \pi \sqrt{3})$ depends on the way multiple scattering is estimated.  

The sum of Eqs. (\ref{SA}) and (\ref{SB}) 
leads to the evolution equation of the bulk velocity of the plasma $\vec{v}_{\mathrm{b}}$ which is 
directly coupled to the photon velocity and to the density contrasts of photons (i.e. $\delta_{\gamma}$) and baryons (i.e. $\delta_{\mathrm{b}}$):
\begin{eqnarray}
&& \partial_{\tau} \vec{v}_{\mathrm{b}} + {\mathcal H} \vec{v}_{\mathrm{b}} = \frac{\vec{J} \times \vec{B}}{a^4 \rho_{\mathrm{b}}} - \vec{\nabla} \phi +  \nu_{\mathrm{b}} \nabla^2 \vec{v}_{\mathrm{b}}+  \frac{4}{3}\frac{\rho_{\gamma}}{\rho_{\mathrm{b}}} \epsilon' (\vec{v}_{\gamma} -\vec{v}_{\mathrm{b}}),
\label{pd1}\\
&& \partial_{\tau} \vec{v}_{\gamma} = - \frac{1}{4} \vec{\nabla} \delta_{\gamma} - \vec{\nabla} \phi +  \nu_{\gamma} \nabla^2 \vec{v}_{\gamma}+ \epsilon' (\vec{v}_{\mathrm{b}} - \vec{v}_{\gamma}),
\label{pd2}\\
&& \partial_{\tau} \delta_{\mathrm{b}} = 3 \partial_{\tau} \psi - \vec{\nabla}\cdot \vec{v}_{\mathrm{b}} + \frac{\vec{J} \cdot \vec{E}}{\rho_{\mathrm{b}} a^4},
\label{pd3}\\
&& \partial_{\tau} \delta_{\gamma} = 4 \partial_{\tau}\psi - \frac{4}{3} \vec{\nabla} \cdot\vec{v}_{\gamma},
\label{pd4}
\end{eqnarray}
where $\nu_{\mathrm{b}}$ and $\nu_{\gamma}$ denote the thermal diffusivity coefficients and 
$\epsilon' = \tilde{n}_{\mathrm{e}} \sigma_{\gamma\mathrm{e}} a$ is the differential optical depth of electron-photon scattering. Equations (\ref{pd1})--(\ref{pd4}) are all consistently written in the longitudinal coordinate system of Eq. (\ref{met1}). 

Dropping the third and fourth terms in Eq.  (\ref{pd5a}) (i.e. the 
thermoelectric and Hall terms)  the evolution equation of the magnetic fields reads, in the one-fluid approximation,  
\begin{equation}
\partial_{\tau} \vec{B} = \vec{\nabla}\times (\vec{v}_{\mathrm{b}} \times \vec{B}) + \lambda  \nabla^2 \vec{B}, \qquad \lambda = \frac{1}{4\pi\sigma}.
\label{pd5}
\end{equation}
 If the Hall and the thermoelectric terms are kept in Eq. (\ref{pd5a}),
Eq. (\ref{pd5}) becomes\footnote{In the present analysis we shall assume a vanishing external gradient in the concentration of the charged species even if this requirement could be relaxed leading, presumably, to an effective evolution of the thermal and magnetic diffusivity coefficients with the physical scale (see, e.g. \cite{kleeorin2,moffat2}). }
\begin{equation}
\partial_{\tau} \vec{B} = \vec{\nabla}\times (\vec{v}_{\mathrm{b}} \times \vec{B}) + \frac{1}{4\pi\sigma} \nabla^2 \vec{B}
+ \vec{\nabla}\times \biggl( \frac{\vec{\nabla} p_{\mathrm{e}}}{e n_{0}} \biggr) - \frac{1}{4\pi e n_{0}}  \vec{\nabla}\times[
(\vec{\nabla} \times \vec{B})\times \vec{B}].
\label{pd5b}
\end{equation}
As in \cite{kaza} we shall assume that the magnetic field is sufficiently small so that the Hall term is negligible (see, however, \cite{kleeorin1}).
We shall also assume that there is no particular alignment between the gradient of the concentration and of the 
electron pressure so that the thermoelectric term is also negligible. 

The standard incompressible closure stipulates that the velocity field is solenoidal but this is not the situation described by Eqs. (\ref{pd1})--(\ref{pd5}) where, in general, $\vec{\nabla}\cdot \vec{v}_{\gamma} \neq 0$ 
and $\vec{\nabla}\cdot \vec{v}_{\mathrm{b}} \neq 0$. Prior to photon decoupling the electron-photon scattering rate
drives the baryon and photon velocities to a common value which will be denoted by $ \vec{v}_{\gamma\mathrm{b}}$ (i.e. $\vec{v}_{\mathrm{b}} \simeq 
\vec{v}_{\gamma} = \vec{v}_{\gamma\mathrm{b}}$). After summing up Eqs. (\ref{pd1}) and (\ref{pd2}) to eliminate the scattering terms we arrive at the following 
equation
\begin{equation}
\partial_{\tau} \vec{v}_{\gamma\mathrm{b}} + \frac{{\mathcal H} R_{\mathrm{b}}}{R_{\mathrm{b}} + 1} \vec{v}_{\gamma\mathrm{b}} = \frac{R_{\mathrm{b}}}{R_{\mathrm{b}} + 1} \frac{\vec{J} \times \vec{B}}{\rho_{\mathrm{b}} a^4} - 
\frac{\vec{\nabla} \delta_{\gamma}}{4 (R_{\mathrm{b}} + 1)} - \vec{\nabla} \phi  + \nu_{\mathrm{th}} \nabla^2 \vec{v}_{\gamma\mathrm{b}},
\label{pd6}
\end{equation}
where $\nu_{\mathrm{th}}$ is the coefficient of thermal diffusivity of the baryon-photon system:
\begin{equation}
\nu_{\mathrm{th}} = \frac{4}{15\,(R_{\mathrm{b}} +1)} \lambda_{\gamma\mathrm{e}}, \qquad \lambda_{\gamma\mathrm{e}} = \frac{a_{0}}{\tilde{n}_{\mathrm{e}} \, \sigma_{\gamma\mathrm{e}} a}, \qquad \sigma_{\gamma\mathrm{e}} = \frac{8}{3}\pi \biggl(\frac{e^2}{m_{\mathrm{e}}^2}\biggr),
\label{pd6a}
\end{equation}
and $R_{\mathrm{b}}(\tau)$ is related to the sound speed of the plasma and it is defined as 
\begin{equation}
R_{\mathrm{b}}(\tau) = \frac{3}{4} \frac{\rho_{\mathrm{b}}}{\rho_{\gamma}} = 0.629 \biggl(\frac{\omega_{\mathrm{b}0}}{0.02258}\biggr) \biggl(\frac{z +1}{1091}\biggr)^{-1},\qquad c_{\mathrm{s b}}(\tau) = \frac{1}{\sqrt{3[R_{\mathrm{b}}(\tau) + 1]}}.
\label{pd6aa}
\end{equation}
In Eq. (\ref{pd3}) $\rho_{\mathrm{b}}$ and $\delta_{\mathrm{b}}$ denote, respectively, the mass density of baryons 
and its inhomogeneities:
\begin{equation}
\rho_{\mathrm{b}} = m_{\mathrm{e}} n_{\mathrm{e}} + m_{\mathrm{i}} n_{\mathrm{i}}, \qquad 
\delta_{\mathrm{b}} = \frac{\delta \rho_{\mathrm{b}}}{\rho_{\mathrm{b}}}.
\label{pd6b}
\end{equation}
In the $\Lambda$CDM case the vector modes 
of the geometry are absent. Strictly speaking the flow considered here is not only compressible (i.e. $\vec{\nabla} \cdot\vec{v}_{\gamma\mathrm{b}} \neq 0$) 
but also irrotational (i.e.  $\vec{\nabla} \times \vec{v}_{\gamma\mathrm{b}} = 0$).
Since $\vec{v}_{\gamma\mathrm{b}}$ is irrotational it can be written as 
$\vec{v}_{\gamma\mathrm{b}} = \vec{\nabla} u_{\gamma\mathrm{b}}$. As an example the 
the equations for $u_{\gamma\mathrm{b}}$ and the magnetic diffusivity equations can be written as:
\begin{eqnarray}
&& \partial_{\tau} u_{\gamma\mathrm{b}} + \frac{{\mathcal H} R_{\mathrm{b}}}{R_{\mathrm{b}} + 1} 
u_{\gamma\mathrm{b}}= 
\frac{ 4 \sigma_{\mathrm{B}} - \Omega_{\mathrm{B}}}{R_{\mathrm{b}} + 1} - \frac{\delta_{\gamma}}{4(R_{\mathrm{b}} + 1)} 
- \phi + \nu_{\mathrm{th}} \nabla^2 u_{\gamma\mathrm{b}} ,
\label{pd8}\\
&& \partial_{\tau} \vec{B} = \vec{\nabla} \times [ \vec{\nabla} u_{\gamma\mathrm{b}} \times \vec{B}] + \lambda \nabla^2 \vec{B},
\label{pd9}
\end{eqnarray}
where, according to the notations and conventions of appendix \ref{APPA} we used that:
\begin{equation}
\vec{J} \times \vec{B} = \frac{4}{3} a^4 \rho_{\gamma} \biggl[ \vec{\nabla} \sigma_{\mathrm{B}} - \frac{1}{4} \vec{\nabla} \Omega_{\mathrm{B}}\biggr].
\label{pd7}
\end{equation}
Equations (\ref{pd8}) and (\ref{pd9}) show that the incompressible closure is not consistent with the presence of scalar inhomogeneities; 
the adiabatic closure seems to be more appropriate and it is at least consistent
with the hydromagnetic equations and with the $\Lambda$CDM paradigm.
The adiabatic closure amounts to requiring that
the fluctuation of the specific entropy (i.e. the ratio between the entropy density 
of the photons and the concentrations of the baryons) vanishes; this condition is equivalent to the requirement that $\delta_{\gamma} = 4 \delta_{\mathrm{b}}/3$. 

 As in the case of the one-fluid hydromagnetic equations in flat space\footnote{In the context of flat-space magnetohydrodynamics the adiabatic closure  amounts to a slightly different requirement, i.e. $\partial_{\tau} [p \rho_{\mathrm{m}}^{-\kappa}]=0$; $\rho_{\mathrm{m}}$ is 
what we called $\rho_{\mathrm{b}}$ (i.e. the mass density of the fluid) and $\kappa$ is the adiabatic index (i.e. the ratio of the heat capacity at constant pressure and constant volume) \cite{krall}.}, different closures 
could be adopted (such as the isothermal closure or the closure with constant Ohmic current). While it could be interesting to explore other closures, the one suggested here is definitively better motivated from the viewpoint of the large-scale initial conditions imposed, at early times, on the Einstein-Boltzmann hierarchy \cite{c1}.  

In the dynamo theory the compressible closure is usually adopted 
(see, e.g. \cite{moffat,parker,zeldovich}) and the same is true 
when the kinetic Reynolds number is very large (see, e.g. \cite{bane}).
However, after electron-positron annihilation $R_{\mathrm{kin}} <1$ while 
$Pr_{\mathrm{magn}} > R_{\mathrm{magn}} \gg 1$ \cite{max3}.
The hypothesis of primeval turbulence (implying the largeness of the kinetic Reynolds number) has been a recurrent theme since the first speculations on the origin of the light nuclear elements. The implications of turbulence for galaxy formation have been pointed out in the fifties by Von Weizs\"aker and Gamow \cite{VW}. They have been scrutinized in the sixties and early seventies by various authors \cite{turb1} (see also \cite{peebles,barrow} and discussions therein). In the eighties it has been argued \cite{hogan} that first-order phase transitions in the early Universe, if present,  can provide a source of kinetic turbulence and, hopefully, the possibility of inverse cascades which could lead to an enhancement of the correlation scale of a putative large-scale magnetic field, as discussed in \cite{olesen1,olesen2,enqvist}. 
The extension of the viewpoint conveyed in the present analysis to earlier times (and larger temperatures) is not implausible but shall not be attempted here. 

\renewcommand{\theequation}{3.\arabic{equation}}
\setcounter{equation}{0}
\section{Predecoupling flow}
\label{sec3}
The solution of the evolution equations (\ref{pd1})--(\ref{pd4}) determines 
the correlation functions of the velocity field and the predecoupling flow. By taking the derivative of Eq. (\ref{pd4}) 
with respect to the conformal time coordinate $\tau$ and by using Eq. (\ref{pd6}) 
a well known second-order equation for $\delta_{\gamma}(k,\tau)$ can be obtained and 
solved with different methods. In particular, within the WKB approximation \cite{peebles1,pavel1,seljak,hu1},
the solution for $\delta_{\gamma}(k,\tau)$ and $u_{\gamma\mathrm{b}}(k,\tau)$ reads, in Fourier space: 
\begin{eqnarray}
&& \delta_{\gamma}(k,\tau) = - \frac{4}{3 c_{\mathrm{sb}}^2} \psi_{\mathrm{m}}(\vec{k}) + 
\sqrt{c_{\mathrm{sb}}} {\mathcal M}_{{\mathcal R}}(k,\tau) \, \cos{[k r_{\mathrm{s}}(\tau)]} e^{- k^2/k_{\mathrm{d}}^2} 
= \frac{4}{3} \delta_{\mathrm{b}}(k,\tau),
\label{IM1}\\
&& u_{\gamma\mathrm{b}}(k,\tau) = -  \frac{1}{k}\overline{{\mathcal M}}_{{\mathcal R}}(\tau) {\mathcal R}_{*}(\vec{k}) 
\sin{[k r_{\mathrm{s}}(\tau)]} \, e^{- k^2/k_{\mathrm{d}}^2},
\label{IM1A}
\end{eqnarray}
where $r_{\mathrm{s}}(\tau)$ is the sound horizon and $k_{\mathrm{d}}(\tau)$ defines the time-dependent scale 
of diffusive damping, i.e.
\begin{equation}
r_{\mathrm{s}}(\tau) = \int_{0}^{\tau}
  c_{\mathrm{sb}}(\tau') \, d\tau' = \int_{0}^{\tau} \frac{d\tau'}{\sqrt{3 [ R_{\mathrm{b}}(\tau') + 1]}},\qquad 
\frac{1}{k_{\mathrm{d}}^2(\tau)} = \frac{2}{5} \int_{0}^{\tau} \lambda_{\gamma\mathrm{e}}(\tau') \, c^2_{\mathrm{sb}}(\tau') \, d\tau'.
\label{IM1B}
\end{equation}
The functions $\psi_{\mathrm{m}}(\vec{k},\tau) = T_{{\mathcal R}}(\tau) {\mathcal R}_{*}(k)$,  
${\mathcal M}_{{\mathcal R}}(k,\tau)$ and  $\overline{{\mathcal M}}_{{\mathcal R}}(\tau)$ 
\begin{equation}
{\mathcal M}_{{\mathcal R}}(k,\tau) = \frac{4}{3^{3/4}}\biggl( \frac{1}{c_{\mathrm{sb}}^2} - 2\biggr) T_{{\mathcal R}}(\tau) {\mathcal R}_{*}(k),\qquad \overline{\mathcal M}_{\mathcal R}(\tau) = \frac{1 + 3 R_{\mathrm{b}}}{\sqrt{3} ( 1 + R_{\mathrm{b}})^{3/4} }  T_{\mathcal R}(\tau),
 \label{IM2}
\end{equation}
are all defined in terms of $T_{{\mathcal R}}(\tau)$ which is a simplified form 
of the transfer function of curvature perturbations discussed, for instance, in appendix B of Ref. \cite{max4}:
\begin{equation}
T_{{\mathcal R}}(\tau)  = 1 - \frac{{\mathcal H}}{a^2} \int_{0}^{\tau} a^2(\tau') d\tau' = 1 - \frac{H}{a} \int_{0}^{t} a(t') dt'.
\label{IM4a}
\end{equation}
Note that, for $\tau \to \tau_{*}$ (where $\tau_{*}$ denotes the last scattering) $T_{{\mathcal R}}(\tau_{*}) \simeq - (3/5)$.
If $k \tau \ll 1$ we have, from Eq. (\ref{IM1}), that $\delta_{\gamma} \to - 8 \psi_{\mathrm{m}}/3$, $c_{\mathrm{sb}} \to 1/\sqrt{3}$ and $\overline{{\mathcal M}}_{{\mathcal R}}(\tau_{*}) \to \sqrt{3}/5$.  
Since the curvature perturbations are distributed as in  Eqs. (\ref{met2})--(\ref{met3}), the correlation function 
of the velocity for unequal times can be written as:
\begin{equation}
\langle v_{i}(\vec{q},\tau) \, v_{j}(\vec{p},\tau^{\,\prime})\rangle = \frac{q_{i} \, q_{j}}{q^2} \,{\mathcal U}(q, |\tau - \tau^{\,\prime}|)\, \delta^{(3)}(\vec{q} + \vec{p}) , 
\qquad 
{\mathcal U}(q, |\tau - \tau^{\,\prime}|)= v(q)  \, \delta(\tau - \tau^{\,\prime}),
\label{IM7A}
\end{equation}
where, to avoid confusions with vector indices, the subscript ``$\gamma\mathrm{b}$"  has been suppressed.
The function $v(q)$ appearing in Eq. (\ref{IM7A}) is
\begin{equation}
v(q) = \tau_{\mathrm {c}} {\mathcal V}(q), \qquad {\mathcal V}(q) = \overline{{\mathcal M}}^2_{{\mathcal R}}(\tau_{*}) \frac{2\pi^2}{q^3}\, {\mathcal P}_{\mathcal R}(q)\, \sin^2{[q r_{\mathrm{s}}(\tau_{*})] } \, e^{ - 2 q^2/q_{\mathrm{d}}^2}.
\label{IM7B}
\end{equation}
The correlation time $\tau_{\mathrm{c}}$ is, by definition, the smallest time-scale when compared with other characteristic times arising in the problem. Because of the exponential suppression of the velocity correlation 
function for $\tau > \tau_{\mathrm{d}}$ (where $\tau_{\mathrm{d}}$ denotes the Silk time \cite{silk}), $\tau_{\mathrm{c}}$ approximately coincides with $\tau_{\mathrm{d}}$. The form of the correlator given in Eq. (\ref{IM7A}) is characteristic of Markovian conducting fluids \cite{kaza,vain0}.
As a consequence of the smallness of $\tau_{\mathrm{c}}$ the velocity correlator 
for unequal times can be approximated with its Markovian form given in Eq. (\ref{IM7B}). 
The velocity field is exponentially suppressed for  $\tau > \tau_{\mathrm{c}} = 1/(k_{\mathrm{max}}^2 \nu_{\mathrm{th}})$ where $\tau_{\mathrm{c}}\simeq  1.95 \times 10^{-6} (z_{*} +1)^2 \, \mathrm{Mpc}$,  $z_{*} \simeq 1090$ and, typically, $k_{\mathrm{max}} \simeq k_{\mathrm{d}}$. Note that $v(q)$ contains $\tau_{\mathrm{c}}$ and it does not have the same dimensions of ${\mathcal V}(q)$.
Defining, for immediate convenience, ${\mathcal P}_{v}(q) = q^3 \,v(q)/(2\pi^2)$, the velocity correlator can be expressed in physical space as 
\begin{eqnarray}
&& \langle v_{i}(\vec{x}, \tau) \, v_{j}(\vec{y},\tau^{\, \prime})\rangle =  \biggl\{ V_{\mathrm{T}}(r) \delta_{ij} + \biggl[V_{\mathrm{L}}(r) 
- V_{\mathrm{T}}(r) \biggr] \frac{r_{i} \, r_{j}}{r^2} \biggr\} \delta(\tau - \tau^{\, \prime}),
\label{IM7C}\\
&& V_{\mathrm{T}}(r) = \int \frac{d q}{q} {\mathcal P}_{v}(q) \biggl[\frac{1}{q^3 r^3} \sin{q r} - \frac{1}{q^2 r^2} \cos{q r} \biggr],
\nonumber\\
&& V_{\mathrm{L}}(r) = \int \frac{d q}{q} {\mathcal P}_{v}(q) \biggl[- \frac{2}{q^3 r^3} \sin{q r} + \frac{2}{q^2 r^2} \cos{q r} + \frac{\sin{q r}}{q r} \biggr],
\label{IM7D}
\end{eqnarray}
where $r = |\vec{x} - \vec{y}|$.
Thanks to the explicit form of the fluid flow, Eqs. (\ref{pd5b}) and (\ref{pd9})  can be solved by neglecting all the terms which are of higher order in the magnetic field 
intensity and by focusing the attention on the coupling 
of the compressible flow to the magnetic field; by following Ref. \cite{kaza,vain0} Eqs. (\ref{pd5b})--(\ref{pd9}) can be solved iteratively as
\begin{eqnarray}
 B_{i}(\vec{k},\tau) &=& \sum_{n =0}^{\infty}  B^{(n)}_{i}(\vec{k},\tau),\qquad {\mathcal G}_{k}(y) = 
 e^{- k^2 \lambda\,y},
\label{NM14aa}\\
B^{(n+1)}_{i}(\vec{k},\tau) &=& \frac{(-i)}{(2\pi)^{3/2}} \, \int_{0}^{\tau} {\mathcal G}_{k}(\tau-\tau_{1}) \, d\tau_{1} \, \int d^{3} q \, \int d^3 p  \,\,\, \delta^{(3)}(\vec{k} - \vec{q} - \vec{p})\,
\nonumber\\
&\times& \epsilon_{m\, n\, i} \, \epsilon_{a\, b\, n} \, (q_{m} + p_{m}) \, v_{a}(\vec{q},\tau_{1}) \, B^{(n)}_{b}(\vec{p},\tau_{1}),
\label{NM14bb}
\end{eqnarray}
where, for simplicity, $\lambda$ is assumed to be constant in time; moreover, as in Eq. (\ref{pd9}),
$v_{a}(\vec{x},\tau) = \partial_{a} u(\vec{x},\tau)$. The magnetic field can then be averaged over the flow by using either Eq. (\ref{IM7A}) or Eq. (\ref{IM7C}) (depending if we work in real space or in Fourier space). 
From Eq. (\ref{NM14bb}) the first few terms of the recursion are $B_{i}^{(0)}(\vec{k},\tau)$, $B_{i}^{(1)}(\vec{k},\tau)$ and $B_{i}^{(2)}(\vec{k},\tau)$. The first term 
is $B_{i}^{(0)}(\vec{k},\tau) = {\mathcal G}_{k}(\tau) B_{i}(\vec{k})$ where $B_{i}(\vec{k})$ parametrizes the initial stochastic magnetic field.
Denoting with  $H_{i}(\vec{k},\tau)$  the magnetic field averaged over the fluid flow, 
the terms containing an odd number of velocities will be zero while the correlators containing an even number of velocities do not vanish
i.e. $ \langle B_{i}^{(2 n + 1)} \rangle =  H_{i}^{(2 n + 1)} =0$  and $\langle B_{i}^{(2 n + 2)} \rangle =  H_{i}^{(2 n + 2)} \neq 0$.
So, for instance, $\langle B_{i}^{(1)}\rangle =0$ while  $\langle B_{i}^{(2)} \rangle =  H_{i}^{(2)}$ is  
\begin{eqnarray}
&& H^{(2)}_{i}(\vec{k},\tau) =  \frac{(-i)^2}{(2\pi)^{3}}\int d^{3} q \, \int d^3 p\, \int d^{3} q^{\,\prime} \, \int d^3 p^{\, \prime} 
\,\,\, \delta^{(3)}(\vec{k} - \vec{q} - \vec{p})\,\,\, \delta^{(3)}(\vec{p} - \vec{q}^{\, \prime} - \vec{p}^{\, \prime})
 \nonumber\\
&&\times \int_{0}^{\tau} d\tau_{1} {\mathcal G}_{k}(\tau-\tau_{1})\, \, \int_{0}^{\tau_{1}} d\tau_{2} \, {\mathcal G}_{p}(\tau_{1}-\tau_{2})\,(q_{m} + p_{m}) \, (q_{m ' }' + p_{m' }') 
  \epsilon_{b\, m' \, n' \,} \, \epsilon_{a'\, b'\, n'} \, \epsilon_{m\, n\, i} \, \epsilon_{a\, b\, n}
 \nonumber\\
 &&\times
  \,  \langle v_{a'}(\vec{q}^{\,\,\prime},\tau_{2}) \, v_{a}(\vec{q},\tau_{1}) \rangle \,B_{b'}(\vec{p}^{\,\prime}).
\label{NM19}
\end{eqnarray}
After averaging the whole series of Eq. (\ref{NM14aa}) term by term the obtained result can be resummed and written as:
\begin{equation}
H_{i}(\vec{k},\tau) = \langle B_{i}^{(0)}(\vec{k},\tau)\rangle  + \langle B_{i}^{(2)}(\vec{k},\tau)\rangle + \langle B_{i}^{(4)}(\vec{k},\tau)\rangle +... =  e^{ - k^2 \, 
\overline{\lambda}\, \tau} B_{i}(\vec{k}).
\label{C21}
\end{equation}
where the magnetic diffusivity 
coefficient $\lambda = 1/(4\pi \sigma) $ inherits a modification stemming from the large-scale flow:
\begin{equation}
\overline{\lambda} = \lambda + v_{0}, \qquad v_{0} = \frac{\tau_{\mathrm{c}}}{3}
\overline{{\mathcal M}}_{{\mathcal R}}^2(\tau_{*})\int\, \frac{d k}{k} \, 
{\mathcal P}_{\mathcal R}(k)\, \sin^2{(k/k_{*})} e^{ - 2 k^2/k_{\mathrm{d}}^2}, 
\label{IM7F}
\end{equation}
and $k_{*} = 1/r_{\mathrm{s}}(\tau_{*})$. The averaging suggested here has been 
explored long ago in the related context of acoustic turbulence  by Vainshtein \cite{vain0}. Prior to decoupling, however, both $\langle v^2 \rangle \propto {\mathcal A}_{\mathcal R} \ll 1$ and  $R_{\mathrm{kin}}\ll 1$. The iterative solution indicated in Eq. (\ref{C21}) seems then to be better defined, in the present case, than in the 
 a kinetically turbulent plasma with strong inhomogeneities. The presence of $v_0$ defines a new diffusion scale associated with 
 the large-scale flow and this scales can be estimated from Eq. (\ref{IM7F}) as
 \begin{equation}
 k_{\mathrm{M}} d_{\mathrm{A}} \simeq \sqrt{\frac{d_{\mathrm{A}}}{\tau_{\mathrm{c}}}} \sqrt{\frac{6 ( 1 - n_{\mathrm{s}})}{{\mathcal A}_{{\mathcal R}} \overline{{\mathcal M}}_{{\mathcal R}}^2(\tau_{*})}} (d_{\mathrm{A}} k_{\mathrm{p}})^{(n-1)/2},
 \label{KM}
\end{equation}
 where subscript M stands for ``Markovian" and
where $d_{\mathrm{A}}$ denotes the (comoving) angular diameter distance to last scattering: since $k_{\mathrm{M}}$ depends on time, it has been evaluated at last scattering.  In the appendix \ref{APPB} the analysis in the non-Markovian approach will be outlined, under some simplifying approximations, 
 with particular attention to the determination of the diffusive scale. 

\renewcommand{\theequation}{4.\arabic{equation}}
\setcounter{equation}{0}
\section{Evolution of the power spectra}
\label{sec4}
Instead of averaging the magnetic field intensities, as explored in the previous section, it is more practical to study directly the evolution equations of the magnetic power spectra with the aim of connecting them to the two-point function of the velocity field. 
In Fourier space the magnetic power spectrum can be defined as
\begin{equation}
 \langle B_{i}(\vec{k},\tau) \, B_{j}(\vec{k}',\tau) \rangle =  M_{ij}(\vec{k},\tau)  \, \delta^{(3)}(\vec{k}+ \vec{k}'), \qquad  M_{ij}(\vec{k},\tau) = P_{ij}(\hat{k}) M(\vec{k},\tau),
 \label{C27}
 \end{equation}
 where $P_{ij}(\hat{k}) = (\delta_{ij} - \hat{k}_{i} \hat{k}_{j})$. The corresponding power spectrum in physical space is 
given by  
\begin{eqnarray}
&& M_{ij}(r,\tau) = M_{\mathrm{T}}(r,\tau) \delta_{ik} + [M_{\mathrm{L}}(r,\tau) - M_{\mathrm{T}}(r,\tau)] \frac{r_{i} r_{k}}{{r^2}},
\label{C28}\\
&& \frac{\partial M_{\mathrm{L}}}{\partial r} + \frac{2}{r}(M_{\mathrm{L}} - M_{\mathrm{T}})=0,
\label{C28A}
\end{eqnarray}
where Eq. (\ref{C28A}) is a consequence of the transversality condition, i.e. $\partial_{i} M^{i j}=0$.
The evolution equation of the magnetic power spectrum in the compressible limit stems from Eq. (\ref{pd9}) and 
can be obtained by taking the conformal time derivative of $M_{ij}(k,\tau)$ as defined in Eq. (\ref{C27}).
The result of this step is the sum of two terms: each of the terms contains the product of the time derivative of the magnetic field with the magnetic field itself for different wavenumbers. The time derivative of the magnetic field 
can be eliminated by means of Eq. (\ref{pd9}) while the magnetic field itself can be written in terms of the integral equation derived always from Eq. (\ref{pd9}). The result of this procedure is, in the compressible case,
\begin{eqnarray}
&& \partial_{\tau} M_{ij} + 2\lambda k^2 M_{ij}  = (-i)^2  \int \frac{d^{3} q}{(2\pi)^{3/2}}\,\int \frac{d^{3} q\,'}{(2\pi)^{3/2}}\,
k_{m} k_{m\,'}\,\, \epsilon_{i m n} \epsilon_{a b n} \epsilon_{j m\,' n\,'} \epsilon_{a\,' b\,' n\,'} \times
\nonumber\\
&& \times \langle v_{a}(\vec{q},\tau) v_{a\,'}(\vec{q}^{\,\prime},\tau\,')\rangle \, M_{b b\,'}(\vec{p},\vec{p}^{\,\prime},\tau, \tau\,'), 
\label{E17}
\end{eqnarray}
where $\vec{p}= (\vec{k} + \vec{q})$ and $\vec{p}^{\,\prime} = (\vec{k} + \vec{q}^{\,\prime})$. Using now the explicit 
form of the velocity correlator  given in Eq. (\ref{IM7A}) and taking the trace of both sides with respect to the 
free tensor indices $i$ and $j$ the following integrodifferential equation is readily obtained 
\begin{equation}
\partial_{\tau} M + 2 k^2 \overline{\lambda} M = \int \frac{d^{3}q}{(2\pi)^3}\,\, v(q) \,\,M(p,\tau) \biggl[ k^2 
- \frac{(\vec{k}\cdot \vec{p})^2}{p^2} + 2 \frac{(\vec{k} \cdot \vec{q}) (\vec{k}\cdot\vec{p}) (\vec{q} \cdot \vec{p}) }{q^2 p^2}
\biggr],
\label{E19}
\end{equation}
where, for convenience, $\vec{p}$ has been kept explicitly but it is in fact $\vec{p} = (\vec{k} - \vec{q})$. Note that $\overline{\lambda}$ is the diffusivity coefficient accounting also for the effects of the flow (see Eq. (\ref{IM7F})).

The same derivation can be performed in the incompressible case but bearing in mind that the correlation function of the velocity field will have a different form which is determined by the transversality of the flow.  The analog of Eq. (\ref{E19}) in the case of the incompressible closure is given by:
\begin{equation}
\partial_{\tau} M + 2 k^2 \overline{\lambda} M = \int \frac{d^{3}q}{(2\pi)^3}\,\, \tilde{v}(q)\,\, M(\vec{p},\tau) \biggl[ k^2 - \frac{(\vec{k}\cdot\vec{q}) (\vec{k}\cdot\vec{p}) (\vec{p}\cdot\vec{q})}{q^2 p^2}\biggr],
\label{D1}
\end{equation}
which coincides with the equation derived by Kazantsev \cite{kaza} except for the presence of $1/(2\pi)^3$ coming from the 
different conventions on the Fourier transforms (we follow here the conventions spelled out explicitly in appendix \ref{APPA}). To correctly 
interpret Eq. (\ref{D1}) it is essential to point out that the velocity field parametrized by $\tilde{v}(q)$ is now solenoidal 
(rather than irrotational as in the case of Eq. (\ref{E19})). To derive Eq. (\ref{D1}) we used the analog of Eq. (\ref{IM7A}) but valid
for the standard incompressible closure, i.e. 
\begin{equation}
\langle \tilde{v}_{i}(\vec{q},\tau) \,\, \tilde{v}_{j}(\vec{p},\tau^{\,\prime})\rangle = \biggl(\delta_{ij} - \frac{q_{i} \, q_{j}}{q^2}\biggr) \,\tilde{{\mathcal U}}(q, |\tau - \tau^{\,\prime}|)\, \delta^{(3)}(\vec{q} + \vec{p}) , 
\qquad 
\tilde{{\mathcal U}}(q, |\tau - \tau^{\,\prime}|)= \tilde{v}(q)  \, \delta(\tau - \tau^{\,\prime}).
\label{D1A}
\end{equation}
The results of Eqs. (\ref{D1}) and (\ref{D1A}) are a consistency check of the whole approach. For sake of comparison, the incompressible equations (derived from Eq. (\ref{D1})) will be sometimes reported after the corresponding results valid in the compressible case. It is finally useful to point out that in Eqs. (\ref{E19}) and (\ref{D1}) the diffusivity coefficient is $\overline{\lambda}$ (and not $\lambda$ itself). This property has been already derived, at the level 
of the field intensities, in Eqs. (\ref{C21})--(\ref{IM7F}). Strictly speaking Eqs. 
(\ref{E19}) and (\ref{D1}) describe the evolution of the magnetic power 
spectra averaged over the fluid flow. In this respect following the notations 
of Kazantsev \cite{kaza} (see also \cite{vain0}) it is sometimes useful to distinguish the ``double" stochastic average $\langle\langle \,\,...\,\,\rangle\rangle$ (over the velocity and over the magnetic field) from the single average (valid either for the velocity or for the magnetic field at the level of the correlators). To preserve
a certain simplicity in the notations the double stochastic average has been avoided.

\subsection{Diffusive approximation}
The integrodifferential equations (\ref{E19}) and (\ref{D1}) can be studied, with two complementary approaches, either in Fourier space or in physical space. By taking the limit $q \to 0$ (while $k$ is held fixed), Eqs. (\ref{E19})--(\ref{D1}) lead to a diffusion equation which is very similar to the one often encountered is the dynamics of wave turbulence 
\cite{sagd1,modpl}. Equations (\ref{E19})--(\ref{D1}) can also be transformed into a Schr\"odinger-like equation in physical space where the analog of the wavefunction is related to the power spectrum introduced in Eq. (\ref{C28}). 
In what follows the relevant evolution equations will be derived in these two complementary approaches. 
The attention will be focused on the compressible mode, however, as a cross-check, the same derivations have been 
also performed with the incompressible closure. In the latter case the equations reproduce 
exactly the results reported in \cite{kaza}. This represents an important cross-check for the 
consistency of the whole procedure. 

In the diffusive limit Eq. (\ref{E19}) is expanded for $\epsilon= |q/k|\ll 1$. The integrand appearing at the right hand side 
of Eq. (\ref{E19}) written in terms of $\epsilon$ becomes
 \begin{eqnarray}
&& M(\vec{p},\tau)\biggl[ k^2 - \frac{(\vec{k}\cdot \vec{p})^2}{p^2} + 2 \frac{(\vec{k} \cdot \vec{q}) (\vec{k}\cdot\vec{p}) (\vec{q} \cdot \vec{p}) }{q^2 p^2}\biggr]
\nonumber\\
&& = k^2 M\biggl( k \sqrt{1 - 2 \epsilon x +\epsilon^2} \biggr) \biggl[ 1 - \frac{(1 - \epsilon x)^2}{1 - 2 \epsilon x + \epsilon^2} + 2 \frac{x ( 1 - \epsilon x) (x - \epsilon)}{1 - 2 \epsilon x +\epsilon^2}\biggr].
\label{E25}
\end{eqnarray}
where $ x = \cos{\vartheta}$ and $\vartheta = (\hat{q}\cdot \hat{k})$. After expanding Eq. (\ref{E25}) 
in powers of $\epsilon$, direct integration over $x$ between $-1$ and $1$ gives
\begin{equation}
\frac{4}{3} k^2 M + \frac{2 k^2}{15} \biggl[ 2 M + 6 k \frac{\partial M}{\partial k} + 
3 k^2 \frac{\partial^2 M}{\partial k^2} \biggr] \epsilon^2.
\label{E26}
\end{equation}
Using the results of Eqs. (\ref{E25}) and (\ref{E26}), Eq. (\ref{E19}) 
\begin{equation}
\frac{\partial M}{\partial \tau} + 2 \lambda k^2 M = \gamma  \biggl[ 2 M + 6 k \frac{\partial M}{\partial k} + 
3 k^2 \frac{\partial^2 M}{\partial k^2} \biggr],\qquad v_{2} = \frac{1}{3!} \int \frac{d^{3} q}{(2\pi)^3} \, q^2 v(q),
\label{E27}
\end{equation}
where $\gamma =2 v_{2}/5$. From the explicit expression of $v(q)$ it is possible to estimate  $\gamma$ 
by replacing $\sin^2 \to 1/2$, $\overline{{\mathcal M}}_{{\mathcal R}}^2(\tau_{*})\to 3/25$, and by fixing the 
upper limit of integration at $k_{\mathrm{d}}$:
\begin{equation}
\gamma =  \frac{ {\mathcal A}_{{\mathcal R}}}{30 (n_{\mathrm{s}} +1)} \overline{{\mathcal M}}^2(\tau_{*})\, \biggl(\frac{k_{\mathrm{d}}}{k_{\mathrm{p}}}\biggr)^{n_{\mathrm{s}} +1} q_{\mathrm{p}}^2 \tau_{\mathrm{c}},\qquad \eta = \gamma \tau,
\label{gam3}
\end{equation} 
where, as already mentioned, $\overline{{\mathcal M}}^2(\tau_{*}) \to 3/25$; the introduction of the variable $\eta$ will be 
useful in section \ref{sec5} since it will simplify the expressions of the Laplace transforms. 
Choosing, for sake of simplicity $k_{\mathrm{d}} \sim 0.1 \, \mathrm{Mpc}^{-1}$ as well as the fiducial set of parameters 
of Eqs. (\ref{met3}) and (\ref{R3a}) $\eta$ can be explicitly estimated; for $\tau \simeq {\mathcal O}(\tau_{*})$ we have that 
$\eta \simeq 6.04 \times 10^{-11}$.

As anticipated after Eqs. (\ref{D1})--(\ref{D1A}), the same kind of equation 
obtained in the compressible limit can be derived when the fluid flow 
is incompressible.  Indeed, from the right hand side of Eq. (\ref{D1}) we have
 \begin{equation}
  M(\vec{p},\tau) \biggl[ k^2 - \frac{(\vec{k}\cdot\vec{q}) (\vec{k}\cdot\vec{p}) (\vec{p}\cdot\vec{q})}{q^2 p^2}\biggr]
  = k^2 M\biggl( k \sqrt{1 - 2 \epsilon x +\epsilon^2} \biggr) \biggl[ 1 - \frac{x ( 1 - \epsilon x) (x - \epsilon)}{1 - 2 \epsilon x +\epsilon^2}\biggr].
\label{C30}
\end{equation}
By expanding Eq. (\ref{C30}) for small $\epsilon$ and by integrating over $x$ between $-1$ and $1$ we have 
\begin{equation}
\frac{4}{3} k^2 M + \frac{2 k^2}{15} \biggl[ 4 M + 2 k \frac{\partial M}{\partial k} + 
k^2 \frac{\partial^2 M}{\partial k^2} \biggr] \epsilon^2,
\label{C31}
\end{equation}
implying, from Eq. (\ref{D1}),
\begin{equation}
\frac{\partial M}{\partial \tau} + 2 \lambda k^2 M = \frac{2}{5} \tilde{v}_{2} \biggl[ 4 M + 2 k \frac{\partial M}{\partial k} + 
k^2 \frac{\partial^2 M}{\partial k^2} \biggr],\qquad \tilde{v}_{2} = \frac{1}{3!} \int \frac{d^{3} q}{(2\pi)^3} \, q^2 \tilde{v}(q).
\label{C32}
\end{equation}
Apart from the slightly different conventions, 
Eq. (\ref{C32}) reproduces the corresponding results of  \cite{kaza} and corroborates the correctness of the results obtained here in 
the compressible case.

\subsection{Schr\"odinger-like equations}
In physical space Eq. (\ref{E19}) can be transformed into a Schr\"odinger-like equation whose 
generalized wavefunction is related to the magnetic power spectrum.
Since, by definition,
\begin{equation}
M(r,\tau) = \frac{1}{(2\pi)^{3/2}} \int d^{3} k e^{- i \vec{k} \cdot \vec{r}} \, M(k, \tau),
\end{equation}
both sides of Eq. (\ref{E19}) can be multiplied by $\exp{[- i\vec{k} \cdot \vec{r}]}$ and then integrated over $d^3 k$. Recalling that $\vec{k} = \vec{q} + \vec{p}$, after simple algebra the evolution of $M(r,\tau)$ can be obtained in physical space:
\begin{equation}
\partial_{\tau} M - 2 \overline{\lambda} \, \frac{\partial^2 M}{\partial r_{a} \partial r^{a}}= 
 -  \frac{\partial^2 ( v \, M )}{\partial r_{a} \partial r^{a}} 
 +  \frac{\partial^2}{\partial r_{i} \, \partial r_{j}}\biggl[ G_{ij} \, v\biggr]- 2  \frac{\partial^2}{\partial r_{i} \, \partial r_{j}} \biggl[G_{ik} F_{kj} \biggr],
 \label{E24}
 \end{equation}
 where it is understood that $M= M(r,\tau)$. According to Eq. (\ref{IM7C}), $v$ coincides 
 with the trace of the correlation function of the velocity in real physical space, i.e. $ v = (2 V_{\mathrm{T}} + 
 V_{\mathrm{L}})$. The tensors $G_{ij}$ and $F_{i j}$ are:
\begin{equation}
G_{ij}= \overline{M} \delta_{ij} + (M - 3 \overline{M}) \frac{r_{i} r_{j}}{r^2}, \qquad 
F_{ij}=  \frac{v - V_{\mathrm{L}}}{2} \delta_{ij} + ( 3 V_{\mathrm{L}} - v) \frac{r_{i} r_{j}}{2 r^2},
\label{E24b}
\end{equation}
where $M(r,\tau)$, $\overline{M}(r,\tau)$ and $M_{\mathrm{L}}(r,\tau)$ are related as follows:
\begin{equation}
\overline{M}(r,\tau) = \frac{M_{\mathrm{L}}(r,\tau)}{2}  = \frac{1}{r^3}\int_{0}^{r} x^2 M(x,\tau) \, d x.
\label{E24c}
\end{equation}
Since for a generic function $f(r,\tau)$ it is easy to show that 
\begin{equation}
\frac{\partial^2 f}{\partial r_{a} \partial r^{a}} = \frac{1}{r^2} \frac{\partial}{\partial r} \biggl[ r^2 \frac{\partial f}{\partial r} \biggr],
\qquad \frac{\partial^2 }{\partial r_{i} \partial r_{j}}[ r_{i} r_{j} f] = \frac{1}{r^2} \frac{\partial^2}{\partial r^2} [ r^4 f].
\label{AUX1}
\end{equation}
Eq. (\ref{E24}) can be transformed as:
\begin{equation}
\frac{\partial \overline{M}}{\partial \tau} + 2 (V_{\mathrm{L}} - \overline{\lambda}) \frac{\partial^2\overline{M}}{\partial r^2} 
+ 2 \biggl[ \frac{\partial V_{\mathrm{L}}}{\partial r} + \frac{4}{r} (V_{\mathrm{L}} - \overline{\lambda}) \biggr] \frac{\partial \overline{M}}{\partial r} 
+ 2\biggl[ \frac{1}{r} \frac{\partial}{\partial r} ( v + V_{\mathrm{L}}) + \frac{1}{r^2} ( 3 V_{\mathrm{L}} - v) \biggr] \overline{M}=0,
\label{SC6}
\end{equation}
which can also be written in the following equivalent form:
\begin{equation}
\frac{\partial \overline{M}}{\partial \tau} + \frac{2}{r^4}  \frac{\partial}{\partial r} \biggl[ r^4 (V_{\mathrm{L}} - \overline{\lambda}) \frac{\partial \overline{M}}{\partial r}\biggr]
+ \frac{2}{r} \biggl[ \frac{\partial}{\partial r} ( v + V_{\mathrm{L}}) + \frac{1}{r} ( 3 V_{\mathrm{L}} - v) \biggr] \overline{M}=0.
\label{SC8}
\end{equation}
By performing the same derivation in the case of Eq. (\ref{D1}),
the analog of Eq. (\ref{SC8}) becomes
\begin{equation}
\frac{\partial \overline{M}}{\partial \tau} + 2 ( \overline{V} -\overline{\lambda}) \frac{\partial^2 \overline{M}}{\partial r^2} 
+ 2(v - 4 \overline{\lambda} + \overline{V}) \frac{1}{r} \frac{\partial \overline{M}}{\partial r} + \frac{2}{r} \overline{M}
\frac{\partial v}{\partial r}=0,
\label{D27}
\end{equation}
where, in this case\footnote{We recall that, in the incompressible case, the explicit forms of the velocity 
correlator in real space is the same as the one given in Eq. (\ref{IM7C}) but with the 
difference that the transversality condition applied to the velocity field implies 
that $\partial_{r} V_{\mathrm{L}} + 2 (V_{\mathrm{L}} - V_{\mathrm{T}})/r=0$; the latter condition leads to
the first expression of Eq. (\ref{D28}).}
\begin{equation}
\overline{V}  =  \frac{1}{r^3} \int_{0}^{r} x^2\, v(x) \, d x= \frac{V_{\mathrm{L}}}{2},\qquad 2 V_{\mathrm{T}} + V_{\mathrm{L}} = v.
\label{D28}
\end{equation}
Equation (\ref{D28}) coincides with the analog equation obtained in \cite{kaza}. It is finally appropriate 
to mention that Eq. (\ref{SC8}) coincides with the equation derived by Vainshtein and Kichatinov in \cite{VK}
once the relevant functions are appropriately renamed. The correspondence 
is such that $V_{\mathrm{T}} \to T_{\mathrm{NN}}$, $V_{\mathrm{L}}\to T_{\mathrm{LL}}$, and $B_{\mathrm{LL}} \to  M_{\mathrm{L}} = 2 \overline{M}$. 
Equations (\ref{SC8}) and (\ref{D1}) can be put in different but equivalent forms like, for instance,
\begin{eqnarray}
&&\frac{\partial\overline{M}}{\partial \tau} + 2 \hat{{\mathcal L}}_{f}(r) \overline{M}(r,\tau) =0,
\label{SE1}\\
&& \frac{\partial\overline{M}}{\partial \tau} + 2 \hat{{\mathcal L}}_{g}(r) \overline{M}(r,\tau) =0,
\label{SE2}
\end{eqnarray}
where Eq. (\ref{SE1}) holds for an incompressible flow while  
Eq. (\ref{SE2}) holds for a compressible flow.
In the incompressible case the linear operator $\hat{{\mathcal L}}_{f}(r)$ takes the form
\begin{equation}
 \hat{{\mathcal L}}_{f}(r) =  - f(r) \frac{\partial^2}{\partial r^2} - f_{1}(r) \frac{\partial}{\partial r} + 
  f_{2}(r),
\label{SE4A}
\end{equation}
where the three functions $f(r)$, $f_{1}(r)$ and $f_{2}(r)$  are defined as 
\begin{equation}
f(r) = \overline{\lambda} - \overline{V}, \qquad f_{1}(r) = \frac{1}{r} [ 4 \overline{\lambda} - \overline{V} - v] 
 ,\qquad f_{2}(r) = \frac{1}{r} \frac{\partial v}{\partial r}.
\label{SE5}
\end{equation}
Conversely, in the compresible case, the linear operator $\hat{{\mathcal L}}_{g}(r)$ is 
\begin{equation}
 \hat{{\mathcal L}}_{g}(r) =  - g(r) \frac{\partial^2}{\partial r^2} - g_{1}(r) \frac{\partial}{\partial r} + 
 g_{2}(r),
\label{SE4}
\end{equation}
and the explicit form of the functions $g(r)$, $g_{1}(r)$ and $g_{2}(r)$ is
 \begin{eqnarray}
 &&  g(r) = \overline{\lambda} - V_{\mathrm{L}}(r),\qquad g_{1}(r) = \frac{\partial}{\partial r} (\overline{\lambda} - V_{\mathrm{L}}) + \frac{4}{r} (\overline{\lambda} - V_{\mathrm{L}}),
\nonumber\\
&& g_{2}(r) = \frac{1}{r} \frac{\partial}{\partial r} (v + V_{\mathrm{L}}) + \frac{1}{r^2} ( 3 V_{\mathrm{L}} - v) = \frac{2}{r} 
\frac{\partial}{\partial r}(V_{\mathrm{T}} + V_{\mathrm{L}}) + \frac{2}{r^2} (V_{\mathrm{L}} - V_{\mathrm{T}}).
\label{SE6}
\end{eqnarray}
By focusing the attention on the compressible case of Eqs. (\ref{SE2}) and (\ref{SE4})--(\ref{SE6}), 
Eq. (\ref{SE2}) can be rewritten by eliminating the first derivative with respect to $r$:
\begin{equation}
\frac{\partial \Psi}{\partial \tau} = 2 g \frac{\partial^2 \Psi}{\partial r^2} - 2 W(r) \Psi,\qquad \overline{M}(r,\tau) = Q(r)\Psi(r,\tau);
\label{SE7}
\end{equation}
the potential $W(r)$ and $Q(r)$ are
\begin{equation}
W(r) = g\biggl[ \frac{1}{2} \biggl(\frac{g_{1}}{g}\biggr)^{\, \prime} + \frac{1}{4}  \biggl(\frac{g_{1}}{g}\biggr)^{2}\biggr] + g_{2}(r), \qquad \frac{Q^{\prime}}{Q} = -\frac{g_{1}}{2 g},
\label{SE8}
\end{equation}
where the prime denotes, in Eqs. (\ref{SE8}) and (\ref{SE9}) a derivation with respect to $r$. Defining, for immediate convenience, 
$g_{1}/g= F^{\prime}/F$, Eq. (\ref{SE8}) implies, together with the definition of $g_{1}(r)$ of Eq. (\ref{SE6}), 
\begin{equation}
W(r) =g(r) \frac{( \sqrt{F})^{\prime\prime}}{\sqrt{F}} + g_{2}(r), \qquad Q(r) = \frac{1}{\sqrt{F(r)}}, \qquad F(r) = r^2 \sqrt{g(r)}.
\label{SE9}
\end{equation}
\renewcommand{\theequation}{5.\arabic{equation}}
\setcounter{equation}{0}
\section{Solutions for the compressible mode}
\label{sec5}
\subsection{Kinetic equations for the power spectra}
It is interesting to solve the equations obtained in the previous section by assuming that the magnetic power spectra are assigned at a specific pivot scale conventionally denoted by $k_{\mathrm{L}}$ \cite{max1}
\begin{equation}
\langle B_{i}(\vec{q},\tau) B_{j}(\vec{p},\tau) \rangle = \frac{2 \pi^2}{q^3} {\mathcal P}_{\mathrm{B}}(k,\tau) P_{ij}(\hat{q}),\qquad P_{ij}(\hat{q})= \delta_{ij} - \hat{q}_{i} \hat{q}_{j},
\label{SD1}
\end{equation}
where $\hat{q}_{i} = q_{i}/q$. 
The evolution equation of ${\mathcal P}_{\mathrm{B}}(k,\tau)$ will be solved by imposing the following boundary conditions 
\begin{equation}
{\mathcal P}(k,0) = A_{\mathrm{B}} \biggl(\frac{k}{k_{\mathrm{L}}}\biggr)^{n_{\mathrm{B}} -1}, \qquad 
{\mathcal P}(k_{\mathrm{L}},\tau) = A_{\mathrm{B}} G(\tau).
\label{SD2}
\end{equation}
Even if $G(\tau)$ will be left unspecified a simple choice would imply
$G(\tau) \propto \delta(\tau)$. The evolution equation for ${\mathcal P}_{\mathrm{B}}(k,\tau)$ stems from Eq. (\ref{E26}) by comparing the parametrization of Eq. (\ref{C27}) with the one of Eq. (\ref{SD1}):
\begin{equation}
\frac{\partial {\mathcal P}_{\mathrm{B}}}{\partial \tau} + 2 k^2 \lambda P_{\mathrm{B}} = \gamma \biggl[ 20\,{\mathcal P}_{\mathrm{B}} - 12\, k \, \frac{\partial {\mathcal P}_{\mathrm{B}}}{\partial k} + 3\, k^2 \, \frac{\partial^2 {\mathcal P}_{\mathrm{B}}}{\partial k^2} \biggr].
\label{SD3}
\end{equation}
Depending on the value of the conductivity there are two regimes which can be identified. In the first 
regime the magnetic diffusivity is subleading (i.e. $k^2 \lambda \ll 10 \gamma$) and Eq. (\ref{SD3}) becomes:
\begin{equation}
\frac{\partial {\mathcal P}_{\mathrm{B}}}{\partial \tau} = \gamma \biggl[ 20\,{\mathcal P}_{\mathrm{B}} - 12\, k \, \frac{\partial {\mathcal P}_{\mathrm{B}}}{\partial k} + 3\, k^2 \, \frac{\partial^2 {\mathcal P}_{\mathrm{B}}}{\partial k^2} \biggr].
\label{SD4}
\end{equation}
When the magnetic diffusivity is subleading Eq. (\ref{SD4}) can be solved 
with the boundary conditions (\ref{SD2}) by standard Laplace transform methods. In particular, 
defining 
\begin{equation}
\overline{{\mathcal P}}_{\mathrm{B}}(k, s) =  \int_{0}^{\infty} e^{- s \tau} \, {\mathcal P}_{\mathrm{B}}(k, \tau)\, d \tau = {\mathcal L} [{\mathcal P}_{\mathrm{B}}(k,\tau)], 
\label{SD5}
\end{equation}
Eq. (\ref{SD4}) together with the first of the two conditions of Eq. (\ref{SD2}) implies the following 
(ordinary) differential equation:
\begin{equation}
\frac{d^2 \overline{{\mathcal P}}_{\mathrm{B}}}{d k^2} - \frac{4}{k} \frac{d \overline{{\mathcal P}}_{\mathrm{B}}}{d k} + 
\biggl[ \frac{20}{3 k^2} - \frac{s}{3 \gamma k^2} \biggr] \overline{{\mathcal P}}_{\mathrm{B}} = - \frac{A_{\mathrm{B}}}{3 \gamma k^2} \biggl(\frac{k}{k_{\mathrm{L}}}\biggr)^{n_{\mathrm{B}}-1}. 
\label{SD6}
\end{equation}
After a standard change of variables Eq. (\ref{SD6}) can be transformed as 
\begin{equation}
\frac{d^2  \overline{Q}_{\mathrm{B}}}{d z^2} - q^2(s,\gamma) \overline{Q}_{\mathrm{B}} = - \frac{A_{\mathrm{B}}}{3 \gamma} e^{p z} 
\qquad q(s,\gamma) = \frac{1}{\sqrt{3 \gamma}} \sqrt{ s - \frac{5}{4} \gamma}, \qquad p = n_{\mathrm{B}} - \frac{7}{2}.
\label{SD8}
\end{equation}
where the variables $z$ and $ \overline{Q}_{\mathrm{B}}(z, s)$ are simply:
\begin{equation}
z = \ln{(k/k_{\mathrm{L}})},\qquad \overline{Q}_{\mathrm{B}}(z, s) = e^{- 5 z/2}  \overline{{\mathcal P}}_{\mathrm{B}}(z,s).
\label{SD7}
\end{equation}
The general solution of Eq. (\ref{SD8}) is
\begin{eqnarray}
\overline{Q}_{\mathrm{B}}(z,s) &=&  \biggl[ c_{1}(s) + \frac{A_{\mathrm{B}}}{6 \gamma q ( p - q(s,\gamma))} \biggr] e^{ q z}
+   \biggl[ c_{2}(s) - \frac{A_{\mathrm{B}}}{6 \gamma q(s,\gamma) ( p + q(s,\gamma))} \biggr] e^{ - q(s,\gamma) z} 
\nonumber\\
&-& \frac{A_{\mathrm{B}}}{3\gamma} \frac{e^{p z}}{p^2 - q^2(s,\gamma)}.
\label{SD9}
\end{eqnarray}
If $z \geq 0$ the second condition of Eq. (\ref{SD2})  implies that  
\begin{eqnarray}
c_{2}(s)  &=& A_{\mathrm{B}}\biggl[ g(s) +  \frac{1}{6 \gamma q(s,\gamma) ( p + q(s,\gamma))} +\frac{1}{3\gamma (p^2 - q^2(s,\gamma))}\biggr],
\nonumber\\
c_{1}(s) &=& - \frac{A_{\mathrm{B}}}{6 \gamma q(s,\gamma) ( p - q(s,\gamma))},
\label{SD10}
\end{eqnarray}
where $g(s)$ simply denotes the Laplace transform of the function $G(\tau)$ which appears in the second relation 
of Eq. (\ref{SD2}). Equation (\ref{SD10}) applies for $z \geq 0$, i.e. 
$k \geq k_{\mathrm{L}}$. Similarly, if $z\leq 0$, the second condition of Eq. (\ref{SD2}) together with the requirement of the existence 
of the inverse Laplace transform implies 
\begin{eqnarray}
c_{1}(s) &=& A_{\mathrm{B}} \biggl[ g(s)  - \frac{1}{6 \gamma q(s,\gamma) ( p - q(s,\gamma))}+  \frac{1}{3\gamma(p^2 - q^2(s,\gamma))}\biggr],
\nonumber\\
 c_{2}(s) &=& \frac{A_{\mathrm{B}}}{6 \gamma q(s,\gamma) ( p + q(s,\gamma))}.
\label{SD11} 
\end{eqnarray}
Equation (\ref{SD11}) applies for $z \leq 0$, i.e. for $k \leq k_{\mathrm{L}}$.
The Laplace antitransform can be easily deduced by appropriately choosing the integration contour in the complex plane. 
In the case $z\leq0$ we have that $\overline{Q}_{\mathrm{B}}(z,s)$ is given by:
\begin{equation}
\overline{Q}_{\mathrm{B}}(z,s) = \frac{A_{\mathrm{B}}}{3 \gamma}\biggl[ \frac{e^{q(s,\gamma) z}}{p^2 - q^2(s,\gamma)} + 3 \,\gamma \, g(s) \, e^{q(s,\gamma) z} 
- \frac{e^{p z}}{p^2 - q^2(s,\gamma)} \biggr].
\label{SD12}
\end{equation}
An analog expression can be deduced in the case $z \geq 0$ by choosing the 
constants of Eq. (\ref{SD9}) as in Eq. (\ref{SD10}). Since the explicit expressions of $q(s,\gamma)$ and $p$ have been given
in Eq. (\ref{SD8}) it is easy to see that the first term in Eq. (\ref{SD12}) has a branch point from the argument of the exponential and a simple pole from the denominator; the second term has only a branch point while the 
third term has just a simple pole.  By taking the inverse Laplace transform of Eq. (\ref{SD12}) the spectrum can be explicitly obtained. The same 
procedure described so far can be repeated in the case $z \geq 0$. The general form of ${\mathcal P}_{\mathrm{B}}(z,\tau)$  is then given by:
\begin{eqnarray}
{\mathcal P}_{\mathrm{B}}(z,\tau) &=& A_{\mathrm{B}}\biggl[ e^{(n_{\mathrm{B}}-1) z}\, e^{5\, \gamma \tau/4 + 3\,\gamma\tau \,(n_{\mathrm{B}} - 7/2)^2} 
- e^{5\, \gamma \tau/4 + 5\, z/2} {\mathcal F}(\tau, |z|, n_{\mathrm{B}})
\nonumber\\
&+& |z| e^{5\, z/2} \int_{0}^{\tau} \frac{G(\tau - u)}{\sqrt{12\,\pi\,\gamma \,u}}\, e^{5\, \gamma u/4 - |z|^2/(12 \gamma u) }\, \frac{d u}{u} 
\biggr],
\label{GEN4}\\
{\mathcal F}(\tau, |z|, n_{\mathrm{B}}) &=& \int_{0}^{\infty}\, e^{ - x\tau}  \frac{ \sin{\biggl( |z| \sqrt{\frac{x}{3\gamma}} \biggr)}}{x + 3 \gamma ( n_{\mathrm{B}} - 7/2)^2} \, d x.
\label{SD13}
\end{eqnarray}
The first and second terms of Eq. (\ref{GEN4}) 
come from the contributions containing, respectively, a simple pole and a branch point in the Laplace transform $\overline{Q}_{\mathrm{B}}(z,s)$; 
the third term in Eq. (\ref{GEN4}) corresponds to the piece containing a simple pole and a branch point. The properties of the solution 
(\ref{GEN4}) are determined by the value of $\gamma\tau$ already 
examined in Eq. (\ref{gam3}).
It is useful to note that when $\gamma \tau$ is kept fixed in such a way that $\gamma \to 0$ and $\tau\to \infty$ Eq. (\ref{SD13}) has a definite 
limit:
\begin{equation}
\lim_{\gamma \to 0, \,\, \tau \to \infty} {\mathcal F}(\tau, |z|, n_{\mathrm{B}})  = \mathrm{Erf}\biggl(\frac{|z|}{\sqrt{ 12 \gamma \tau}}\biggr).
\label{SD15}
\end{equation}

As already stressed in connection with Eqs. (\ref{SD3})--(\ref{SD4}),
the solution (\ref{GEN4}) holds in the limit $k^2 \lambda < 10 \gamma$. In the opposite limit (i.e. $k^2 \lambda > 10 \gamma$) it is possible to look for the 
solution by bootstrapping Eq. (\ref{GEN4}):
\begin{equation}
{\mathcal P}_{\mathrm{B}}(k,\tau) = e^{5\, \gamma\tau/4 + 3 \, (n_{\mathrm{B}} - 7/2)^2 \, \gamma \tau} \,\,{\mathcal G}(k). 
\label{SD16}
\end{equation}
Inserting Eq. (\ref{SD16}) into Eq. (\ref{SD4}) and eliminating the first derivatives with respect to $k$ the following equation can be obtained in terms of the rescaled function $\overline{{\mathcal G}}_{k} = {\mathcal G}(k)/k^2$:
\begin{equation}
\frac{d^2 \overline{{\mathcal G}}}{d k^2} + \biggl[ - \frac{2 \lambda}{3 \gamma} - \frac{\nu^2 - 1/4}{k^2}\biggr] \overline{{\mathcal G}} =0, \qquad \nu= | n_{\mathrm{B}} - 7/2|.
\label{SD17}
\end{equation}
The general solution of Eq. (\ref{SD17}) is given in terms of a linear combination of modified Bessel functions $I_{\nu}(z)$ and $K_{\nu}(\nu)$ \cite{ABR}.
By assuming that the initial magnetic power spectrum is such that $n_{\mathrm{B}}<7/2$ the solution of Eq. (\ref{SD17}) is:
\begin{equation}
{\mathcal G}(k) = \frac{2^{1 - \nu} A_{\mathrm{B}}}{\Gamma(\nu)} \biggl(\frac{k}{k_{\mathrm{L}}}\biggr)^{5/2} \biggl(\frac{k_{\gamma}}{k_{\mathrm{L}}} \biggr)^{ - \nu}
\, \, K_{\nu}(k/k_{\gamma}), \qquad k_{\gamma} = \sqrt{\frac{3\gamma}{2 \lambda}}, \qquad \nu = \frac{7}{2} - n_{\mathrm{B}},
\label{SD18}
\end{equation}
which satisfies the correct boundary conditions since, in the limit $k\ll k_{\gamma}$, we have that ${\mathcal G}(k) \to A_{\mathrm{B}} (k/k_{\mathrm{L}})^{n_{\mathrm{B}} - 1}$.  The results obtained so far partially challenges the standard argument leading to the calculation of the magnetic diffusivity scale which is 
phenomenologically important insofar as it determines the scale of the exponential suppression of the magnetic power 
spectrum. Prior to decoupling the magnetic diffusivity scale is estimated by requiring that $k^2_{\sigma} \simeq 4 \pi \sigma {\mathcal H}$. The latter condition stems directly from the magnetic diffusivity equation and it 
totally neglects the flow.

The solution (\ref{GEN4}) can be used to compute the evolution of the spectral index induced by the predecoupling 
flow. The spectral index of the magnetic field is defined as 
\begin{equation}
N - 1 = \frac{\partial \ln{{\mathcal P}_{\mathrm{B}}}}{\partial \ln{k}} \equiv \frac{1}{{\mathcal P}_{\mathrm{B}}} 
\frac{\partial{\mathcal P}_{\mathrm{B}} }{\partial z}.
\label{sin23}
\end{equation}
Having assigned the spectrum at a fiducial  (pivot) scale, it is interesting to ask what happens 
to the power spectrum for larger length-scales (i.e.  smaller wavenumbers). 
From Eq. (\ref{GEN4}) it is possible to derive an evolution equation for $N$ as it is defined in Eq. (\ref{sin23}). After 
simple algebra the result is given by:
\begin{equation}
\frac{\partial N}{\partial \tau} = \gamma \biggl[ - 15 \frac{\partial N}{\partial z} + 3 \frac{\partial^2 N}{\partial z^2} + 
6 N \biggl(\frac{\partial N}{\partial z} \biggr) \biggr].
\label{sin24}
\end{equation}
The solution of Eq. (\ref{sin24}) can be found in the form $N(z,\tau) = n_{\mathrm{B}} -1 + \delta N(z,\tau)$ 
with initial conditions $\delta N(z,0) =0$. In this case Eq. (\ref{sin24}) becomes:
\begin{equation}
\frac{1}{\gamma} \frac{\partial \delta N}{\partial \tau} = 3 \frac{\partial^2 \delta N}{\partial z^2} + 
( 6 n_{\mathrm{B}} -21) \frac{\partial \delta N}{\partial z}.
\label{sin25}
\end{equation} 
Using the standard Laplace transform technique we have that 
\begin{equation}
N(z, \tau) = n_{\mathrm{B}} + \frac{|z|}{12\pi \gamma^3 \tau^3} e^{- y^2(z,\tau)},\qquad y(z,\tau) = \frac{1}{\sqrt{12 \gamma\tau}}\biggl[z + 6 \gamma\tau \biggl(n_{\mathrm{B}} - \frac{7}{2}\biggr)\biggr]^2.
\label{sin26}
\end{equation}
The result of Eq. (\ref{sin26}) matches pretty well with the result obtained directly 
from Eq. (\ref{sin23}) by using the exact solution of Eq. (\ref{GEN4}), as it can be 
seen by plotting the respective functions for specific values of the spectral index. 
To avoid lengthy digression this analysis will be omitted. 

\subsection{Evolution in physical space}
The Schr\"odinger-like equation in the compressible case (see Eqs. (\ref{SE7}), (\ref{SE8}) and (\ref{SE9}))
allows for a semi-quantitative description of the limits $r\to 0$ (where magnetic 
diffusivity dominates) and $r\to \infty$ (where the large-scale flow dominates). 
To deduce the potential in the limit $r\to 0$  the correlation 
functions of the velocity can be expanded in the limit  $R = r k_{*}  < 1$ 
where $k_{*}$ is an auxiliary scale in the space of the wavenumbers. 
Let us therefore start with $V_{\mathrm{T}}(r)$ and $V_{\mathrm{L}}(r)$ written as
\begin{eqnarray}
V_{\mathrm{T}}(r) &=& \overline{{\mathcal M}}_{{\mathcal R}}^2(\tau_{*}) {\mathcal A}_{\mathcal R} \biggl(\frac{k_{0}}{k_{\mathrm{p}}}\biggr)^{n_{\mathrm{s}} -1} \, \int_{k_{0}/k_{\mathrm{p}}}^{\infty} \frac{d x}{x} 
x^{n -1} \sin^2{( x \alpha)}\, e^{ - 2 x^2 \beta} \, A(x, R),
\nonumber\\
V_{\mathrm{L}}(r) &=& \overline{{\mathcal M}}_{{\mathcal R}}^2(\tau_{*}) {\mathcal A}_{\mathcal R} \biggl(\frac{k_{0}}{k_{\mathrm{p}}}\biggr)^{n_{\mathrm{s}} -1} \, \int_{k_{0}/k_{\mathrm{p}}}^{\infty} \frac{d x}{x} 
x^{n -1} \sin^2{( x \alpha)} e^{ - 2 x^2 \beta} \, B(x, R),
\label{IM32}
\end{eqnarray}
where $k_{0}$ can be estimated from the inverse of the comoving angular diameter distance 
to last scattering. In Eq. (\ref{IM32}) the functions $A(x, R)$ and $ B(x, R)$
are defined as
\begin{eqnarray}
A(x, R) &=& \biggl[ \frac{1}{x^3 R^3} \sin{x R} - \frac{1}{x^2 R^2} \cos{x R} \biggr],
\nonumber\\
B(x, R) &=& \biggl[ -\frac{2}{x^3 R^3} \sin{x R} + \frac{2}{x^2 R^2} \cos{x R} + \frac{\sin{x R}}{x R} \biggr].
\label{IM33}
\end{eqnarray}
In Eq. (\ref{IM32}) the following rescaled quantities have been introduced:
\begin{equation}
\alpha = r_{\mathrm{s}}(\tau_{*}) k_{*}, \qquad \beta = \frac{k_{*}^2}{k_{\mathrm{d}}^2},\qquad x = \frac{k}{k_{*}}, \qquad R = r k_{*}.
\label{IM34}
\end{equation}
The expansion of $V_{\mathrm{T}}(R)$ and $V_{\mathrm{L}}(R)$ for $R<1$ 
becomes therefore
\begin{eqnarray}
&&V_{\mathrm{T}}(R) = {\mathcal C}(n_{\mathrm{s}}, {\mathcal A}_{\mathcal R})\biggl[ {\mathcal I}(n_{\mathrm{s}}, \alpha,\beta) -{\mathcal I}(n_{\mathrm{s}}+2, \alpha,\beta)  \frac{R^2}{10}
+ {\mathcal I}(n_{\mathrm{s}}+4, \alpha,\beta) \frac{R^4}{280}+...\biggr],
\label{IM35}\\
&&V_{\mathrm{L}}(R) = {\mathcal C}(n_{\mathrm{s}}, {\mathcal A}_{\mathcal R})\biggl[ {\mathcal I}(n_{\mathrm{s}}, \alpha,\beta) - {\mathcal I}(n_{\mathrm{s}}+2, \alpha,\beta) \frac{3 R^2}{10} +  {\mathcal I}(n_{\mathrm{s}}+4, \alpha,\beta) \frac{R^4}{56}+...\biggr],
\label{IM36}
\end{eqnarray}
where 
\begin{equation}
 {\mathcal C}(n_{\mathrm{s}}, {\mathcal A}_{\mathcal R}) = \tau_{\mathrm{c}} \frac{{\mathcal A}_{\mathcal R}}{25} \biggl(\frac{k_{*}}{k_{\mathrm{p}}}\biggr)^{n -1}, \qquad {\mathcal I}(n_{\mathrm{s}}, \alpha,\beta) = \int_{0}^{\infty} \frac{d x}{x}\, x^{n_{\mathrm{s}} -1} \, 
\sin^2{(x \alpha)} \, e^{ - 2 x^2 \beta}.
\label{IM37}
\end{equation}
To discuss in a semi-quantitative manner the properties of Eq. (\ref{SE7}) it is useful (even if not necessary) to identify 
 $k_{*} \simeq \pi/d_{\mathrm{A}}(\tau_{*})$. Following the approach of Zeldovich (see, e.g. \cite{zeldovich}) it is useful 
 to introduce an interpolating form of the correlation function of the velocity which is 
 suppressed for $r \to \infty$ and  leads to the correct limits of Eqs. (\ref{IM35}) and (\ref{IM36}) for $r\to 0$.
The interpolating form is given by 
\begin{equation}
V_{\mathrm{L}}(r) = v_{0} e^{- 3 \mu r^2},\qquad V_{\mathrm{T}}(r) = v_{0} e^{- \mu r^2},
\label{IM38}
\end{equation}
where $v_{0}$ has been already introduced in Eq. (\ref{IM7F}) and coincides with the normalization which can be 
derived from Eqs. (\ref{IM35})--(\ref{IM37}) in the limit $\overline{{\mathcal M}}_{{\mathcal R}}(\tau_{*}) \to 
\sqrt{3}/5$. Using Eq. (\ref{IM38}) the effective potential of Eqs. (\ref{SE8})--(\ref{SE9})  can be computed. For instance 
$g(r) = v_0[ 1 + \varepsilon - \exp{(- 3 \mu r^2)}]$ where $\varepsilon= \lambda/v_{0} \ll 1$. 
In the limits $r\to 0$ and $r\to\infty$ the potential $W(r)$ becomes
\begin{equation}
\lim_{r\sqrt{\mu} \to 0}  W(r) = \frac{2\lambda}{r^2}, \qquad \lim_{r\sqrt{\mu}  > 1 }  W(r) = \frac{2 \overline{\lambda}}{r^2},
\label{IM40}
\end{equation}
showing that for intermediate and large scales the suppression of the correlation function of the magnetic field is 
determined by the large-scale flow contained in $\overline{\lambda}$. 
If $v_{0} \gg \lambda$ the potential can become negative. Equation (\ref{SE7}) may have negative ``energy" levels, i.e. 
solutions for $\Psi(r,\tau)\simeq \exp{[- 2 E\tau]} \Phi(r)$ which are exponentially increasing in time (i.e. $E<0$). 
For at least one level to exist we should have, roughly, that
\begin{equation}
\int_{r_{1}}^{r_{2}}\frac{\sqrt{E - W(r)}}{\sqrt{g(r)}} dr \geq \frac{\pi}{2},
\label{IM41}
\end{equation}
where $r_{1}$ and $r_{2}$ are the inversion points.  Equation (\ref{IM41}) can be verified but
the growth rate of the field is $\Gamma \sim \mu v_{0}$ leads to a negligible 
integrated growth at last scattering, i.e.  $\Gamma \tau_{*} \sim {\mathcal O}(10^{-14})$ even assuming $\sqrt{\mu} \sim 1/d_{\mathrm{A}}(\tau_{*})$. This semi-quantitative argument (also employed in the case 
of gyrotropic turbulence \cite{zeldovich}) follows from the analogy with the Schr\"odinger 
equation in the WKB approximation.
\renewcommand{\theequation}{6.\arabic{equation}}
\setcounter{equation}{0}
\section{Concluding remarks}
\label{sec6}
The standard treatment adopted for the evolution of predecoupling magnetic fields 
usually neglects two aspects which are the starting point of the present investigation, namely the compressibility of the plasma and the large-scale flow induced by curvature perturbations. A third related coincidence is that, prior to decoupling, the magnetic Reynolds number 
is roughly $20$ orders of magnitude larger than the kinetic Reynolds number which is, in turn, smaller than one.
The basic idea has been to derive an integrodifferential equation valid for the compressible mode and for a standard adiabatic closure. In Fourier space, the diffusive approximation leads to a nonlocal 
diffusion equation similar to the ones often discussed in wave turbulence. In physical space 
the integrodifferential equation can be transformed into a Schr\"odinger-like equation 
whose effective potential depends on the spectral properties of the large-scale flow.  Some applications have been discussed by solving the corresponding equations: they range 
from the effective evolution of the magnetic spectral index to the qualitative 
discussion of the spectrum of the eigenvalues of the equation for the magnetic power spectra in physical space. 

In the absence of large-scale flow the magnetic power spectra are suppressed at small scales (i.e. large wavenumbers) because of the finite value of the conductivity which implies an effective (comoving) wavenumber $k_{\sigma}$
\begin{equation}
k_{\sigma} \simeq 2.5 \times 10^{10} \,\biggl(\frac{d_{\mathrm{A}}}{14116\,\, \mathrm{Mpc}}\biggr)^{-1/2} \,\,\mathrm{Mpc}^{-1},
\label{cr1}
\end{equation}
where $d_{\mathrm{A}} \simeq 14116 \,\, \mathrm{Mpc}$ denotes the (comoving) angular diameter distance to last scattering. In units of $d_{\mathrm{A}}$ we have that $k_{\sigma} \, d_{\mathrm{A}}\simeq {\mathcal O}(10^{14})$ implying  that the corresponding length-scale is much smaller than the Hubble radius 
at last scattering, as expected. The results of the present paper suggest, however, that 
the correct magnetic diffusivity length-scale is much larger than $k_{\sigma}^{-1}$.  
Using the Markovian approximation it has been shown that 
the magnetic field can be averaged over the large-scale flow and the resulting diffusive scale can be estimated from Eq. (\ref{KM}) and it is
\begin{equation}
k_{\mathrm{M}} \simeq 1.42 \times 10^{3}\,\,\biggl(\frac{{\mathcal A}_{\mathcal R}}{2.43\times 10^{-9}}\biggr)^{-1/2}\,\,\, \mathrm{Mpc}^{-1},
\label{cr2}
\end{equation}
 where the fiducial set of parameters of Eq. (\ref{R3a}) has been assumed in the context 
 of the vanilla $\Lambda$CDM scenario. Not only $k_{\mathrm{M}} \ll k_{\sigma}$ but, in units of 
$d_{\mathrm{A}}$, it turns out  that  $k_{\mathrm{M}} d_{\mathrm{A}} \simeq {\mathcal O}(10^{7}) d_{\mathrm{A}}$.
From the analysis of the evolution equation of the power spectrum in the diffusive approximation
 it is possible to derive yet another dissipative scale, i.e. $k_{\gamma}$ (see Eq. (\ref{SD18}) for a definition) 
whose explicit value, always in terms of our fiducial set of parameters, is:
\begin{equation}
k_{\gamma} \simeq 2.16 \times 10^{6}\,\,\biggl(\frac{{\mathcal A}_{\mathcal R}}{2.43\times 10^{-9}}\biggr)^{1/2}\,\,\, \mathrm{Mpc}^{-1}.
\label{cr3}
\end{equation}
By comparing Eqs. (\ref{cr1}), (\ref{cr2}) and (\ref{cr3}) it is clear that the following (approximate) hierarchy holds, i.e. 
$k_{\mathrm{M}} < k_{\gamma} < k_{\sigma}$.  Both the results for $k_{\mathrm{M}}$ and $k_{\gamma}$ have been 
derived, directly or indirectly, by assuming the Markovian approximation for the velocity correlator. 
If the Markovian approximation is relaxed the diffusive wavenumber does not get larger but even 
smaller. From Eq. (\ref{NM28}) it is possible to obtain that
\begin{equation}
k_{\mathrm{nM}} \simeq 7.5\times 10^{-2} \,\,\biggl(\frac{{\mathcal A}_{\mathcal R}}{2.43\times 10^{-9}}\biggr)^{1/2}\,\,\, \mathrm{Mpc}^{-1}.
\label{cr4}
\end{equation}
In summary, it has been shown that the large-scale flow can affect the evolution of the magnetic power 
spectra at large scales not only by potentially shifting the effective spectral index but also 
by changing the diffusive scales. Thanks to the results derived in the present paper, the  evolution of the magnetic 
power spectra prior to decoupling can be addressed in terms of a novel set of equations which take into account the effects of the large-scale flow and which can be explicitly solved in various physical limits.
According to the present results it does not seem correct to treat the predecoupling plasma 
by simply assuming an incompressible flow; the latter closure is sound in the absence of large-scale curvature perturbations, in flat space-time and for high (kinetic and magnetic) Reynolds numbers. None of these three 
assumptions are verified after electron-positron annihilation and prior to last scattering. A closer scrutiny of the description developed here seems therefore both motivated and potentially rewarding. 
\section*{Acknowledgments}
It is a pleasure to thank T. Basaglia and the CERN scientific information service for the kind assistance. 
\newpage
\begin{appendix}
\renewcommand{\theequation}{A.\arabic{equation}}
\setcounter{equation}{0}
\section{Basic conventions}
\label{APPA}
In this appendix some basic conventions will be summarized. As an example, the Fourier transform of the 
magnetic field and of the velocity field are defined as
\begin{equation}
B_{i}(\vec{x},\tau) = \int \frac{d^{3} q}{(2\pi)^{3/2}} \, B_{i}(\vec{q},\tau) \, e^{- i \vec{q}\cdot \vec{x}},\qquad v_{i}(\vec{x},\tau) = \int \frac{d^{3} q}{(2\pi)^{3/2}} \, v_{i}(\vec{q},\tau) \, e^{- i \vec{q}\cdot \vec{x}}.
\label{FT1}
\end{equation}
While the magnetic field $\vec{B}$ is strictly divergenceless (i.e. $\vec{\nabla}\cdot \vec{B}=0$), the velocity 
field may be either solenoidal (i.e. $\vec{\nabla}\cdot\vec{v}=0$) or not solenoidal (i.e. $\vec{\nabla}\cdot\vec{v}\neq 0$). 
If the flow is strictly solenoidal it is also incompressible. In the vanilla $\Lambda$CDM case the flow 
is irrotational and compressible. The following definitions will also be employed in section \ref{sec2}:
\begin{equation}
\Omega_{\mathrm{B}}(\vec{x},\tau) =  \int \frac{d^{3}k}{(2\pi)^{3/2}}\, \Omega_{\mathrm{B}}(\vec{k},\tau) \, e^{- i \vec{k}\cdot \vec{x}},\qquad 
\Pi_{ij}(\vec{x},\tau) = \int  \frac{d^{3}k}{(2\pi)^{3/2}} \, \Pi_{ij}(\vec{k},\tau) \, e^{- i \vec{k}\cdot\vec{x}},
\label{FT2}
\end{equation}
where $\Omega_{\mathrm{B}}(\vec{k},\tau)$ and $\Pi_{ij}(\vec{k},\tau)$ are given by:
\begin{eqnarray}
\Omega_{\mathrm{B}}(\vec{k},\tau) &=& \frac{1}{8\pi a^4 \rho_{\gamma}} \int \,\frac{d^{3}q}{(2\pi)^{3/2}} \, B_{k}(\vec{q},\tau) 
B^{k}(\vec{k}- \vec{q}, \tau),
\label{FT4}\\
\Pi_{ij}(\vec{k},\tau) &=& \frac{1}{4\pi a^4} \int \,\frac{d^{3}q}{(2\pi)^{3/2}} \, \biggl[ B_{i}(\vec{q},\tau) B_{j}(\vec{k}-\vec{q},\tau) - \frac{1}{3} B_{k}(\vec{q},\tau) B^{k}(\vec{k} - \vec{q},\tau) \delta_{ij}\biggr].
\label{FT5}
\end{eqnarray}
Recalling that $(p_{\gamma} + \rho_{\gamma}) \nabla^2 \sigma_{\mathrm{B}} = \partial_{i} \partial_{j} \Pi^{ij}_{\mathrm{B}}$
we also have, in Fourier space, that  
\begin{equation}
\sigma_{\mathrm{B}}(\vec{k},\tau)  = \frac{3\, \hat{k}_{i} \,\hat{k}_{j} }{16\pi a^4 \rho_{\gamma}} 
\int\, \frac{d^{3} q}{(2\pi)^{3/2}}\biggl[ B^{i}(\vec{q},\tau) B^{j}(\vec{k}-\vec{q},\tau) - \frac{\delta^{ij}}{3} B_{k}(\vec{q},\tau) 
B^{k}(\vec{k} - \vec{q},\tau) \biggr].
\label{FT8}
\end{equation}
Finally Eqs. (\ref{FT2}) and (\ref{FT8}) imply
\begin{equation}
\sigma_{\mathrm{B}}(\vec{k},\tau) + \frac{1}{2} \Omega_{\mathrm{B}}(\vec{k},\tau) = 
\frac{3\, \hat{k}_{i}\, \hat{k}_{j}}{16\pi a^4 \rho_{\gamma}} \int \frac{d^{3} q}{(2\pi)^{3/2}} B^{i}(\vec{q},\tau) B^{j}(\vec{k} - 
\vec{q},\tau).
\label{FT9}
\end{equation}
\renewcommand{\theequation}{B.\arabic{equation}}
\setcounter{equation}{0}
\section{Non Markovian approach}
\label{APPB}
In the non-Markovian approximation the correlator of the velocities in the compressible 
case is still Gaussian and it is proportional to the correlator of the curvature perturbations:
\begin{equation}
\langle v_{i}(\vec{q},\tau_{1}) v_{j}(\vec{p},\tau_{2}) \rangle = \frac{q_{i} q_{j}}{q^2} \Gamma(q,\tau_{1},\tau_{2}) 
\langle {\mathcal R}_{*}(\vec{q}) {\mathcal R}_{*}(\vec{p}) \rangle,
\label{NM6}
\end{equation}
having introduced the function $\Gamma(q, \tau_{1},\tau_{2})$
\begin{equation}
\Gamma(q, \tau_{1},\tau_{2}) = \overline{\mathcal M}_{\mathcal R}(\tau_{1}) \, 
\overline{\mathcal M}_{\mathcal R}(\tau_{2})\, \sin{(q c_{\mathrm{sb}} \tau_{1})} \, \sin{(q c_{\mathrm{sb}} \tau_{2})}\, 
e^{- q^2 \nu_{\mathrm{th}} (\tau_{1} + \tau_{2})},
\label{NM7}
\end{equation}
where $ \overline{\mathcal M}_{\mathcal R}(\tau)$ is given by
\begin{equation}
 \overline{\mathcal M}_{\mathcal R}(\tau) = \frac{1 + 3 R_{\mathrm{b}}}{\sqrt{3} ( 1 + R_{\mathrm{b}})^{3/4} } {\mathcal T}_{\mathcal R}(\tau) \to \frac{\sqrt{3}}{5}.
\label{NM8}
\end{equation}
The limit in the second expression holds after matter-radiation equality and for $R_{\mathrm{b}}\to 0$. 
As explained this approximation is justified at last scattering where $R_{\mathrm{b}} \sim 0.6$.
 The expression of Eq. (\ref{NM7}) can also be written as:
 \begin{equation}
 \Gamma(q, \tau_{1},\tau_{2}) =\frac{1}{2} \biggl\{ \cos{[q c_{\mathrm{sb}} (\tau_{1} - \tau_{2})]} - 
 \cos{[q c_{\mathrm{sb}} (\tau_{1} + \tau_{2})]}\biggr\} e^{- q^2 \nu_{\mathrm{th}} ( \tau_{1} + \tau_{2})}  \overline{\mathcal M}_{\mathcal R}(\tau_{1}) \, 
\overline{\mathcal M}_{\mathcal R}(\tau_{2}).
\label{NM9}
\end{equation}
With these notations the correlator of the velocities can be written as:
\begin{equation}
\langle v_{i}(\vec{q},\tau_{1}) v_{j}(\vec{p},\tau_{2}) \rangle = \frac{q_{i} q_{j}}{q^2} \, v(q) \, \Gamma(q,\tau_{1},\tau_{2}) 
\delta^{(3)}(\vec{q} + \vec{p}), \qquad v(q) = \frac{2 \pi^2}{q^3} {\mathcal P}_{{\mathcal R}}(q).
\label{NM10}
\end{equation}
To simplify the problem it is practical to use the limit $\sigma \to \infty$ where the magnetic diffusivity 
equation is given by:
\begin{equation}
\frac{\partial \vec{B}}{\partial \tau} = \vec{\nabla}\times (\vec{v}_{\gamma\mathrm{b}} \times \vec{B}).
\label{NM11}
\end{equation}
By writing Eq. (\ref{NM11})  in Fourier space we have:
\begin{equation}
\partial_{\tau} B_{i}= \frac{(-i)}{(2\pi)^{3/2}} \int d^{3} q\, \int d^{3} p \, \delta^{(3)}(\vec{k} - \vec{q} - \vec{p}) \, 
 \epsilon_{m\, n\, i} \, \epsilon_{a\, b\, n} \, (q_{m} + p_{m}) \, v_{a}(\vec{q},\tau) \, B_{b}(\vec{p},\tau), 
\label{NM12}
\end{equation}
The solution of Eq. (\ref{NM12}) can be formally written as:
\begin{equation}
B_{i}(\vec{k},\tau) = \frac{(-i)}{(2\pi)^{3/2}} \, \int_{0}^{\tau} \, d\tau_{1} \, \int d^{3} q \, \int d^3 p  \, \delta^{(3)}(\vec{k} - \vec{q} - \vec{p})\, 
\epsilon_{m\, n\, i} \, \epsilon_{a\, b\, n} \, (q_{m} + p_{m}) \, v_{a}(\vec{q},\tau_{1}) \, B_{b}(\vec{p},\tau_{1}).
\label{NM13}
\end{equation}
Equation (\ref{NM13}) can be solved by iteration as 
\begin{eqnarray}
 B_{i}(\vec{k},\tau) &=& \sum_{n =0}^{\infty}  B^{(n)}_{i}(\vec{k},\tau),
\nonumber\\
B^{(n+1)}_{i}(\vec{k},\tau) &=& \frac{(-i)}{(2\pi)^{3/2}} \, \int_{0}^{\tau} \, d\tau_{1} \, \int d^{3} q \, \int d^3 p  \,\,\, \delta^{(3)}(\vec{k} - \vec{q} - \vec{p})\,
\nonumber\\
&\times& \epsilon_{m\, n\, i} \, \epsilon_{a\, b\, n} \, (q_{m} + p_{m}) \, v_{a}(\vec{q},\tau_{1}) \, B^{(n)}_{b}(\vec{p},\tau_{1}).
\label{NM14}
\end{eqnarray}
We can then average the magnetic field over the velocity field. The terms containing an odd number 
of velocities will be zero while the correlators containing an even number of velocities do not vanish, in formulas
\begin{equation}
 \langle B_{i}^{(2 n + 1)} \rangle =  H_{i}^{(2 n + 1)} =0,\qquad \langle B_{i}^{(2 n + 2)} \rangle =  H_{i}^{(2 n + 2)} \neq 0.
\label{NM15}
\end{equation}
This conclusion holds both in the Markovian and in the non-Markovian case since it is directly related to the 
Gaussianity of the curvature perturbations.
Let us now write explicitly the contributions up to second order and let us compute the first few terms in the expansion:
\begin{eqnarray}
B_{i}^{(2)}(\vec{k},\tau)&=& \frac{(-i)}{(2\pi)^{3/2}} \int d^{3} q \, \int d^3 p  \,\,\, \delta^{(3)}(\vec{k} - \vec{q} - \vec{p})\,
\nonumber\\
&\times& \epsilon_{m\, n\, i} \, \epsilon_{a\, b\, n} \, (q_{m} + p_{m}) \int_{0}^{\tau} \, d\tau_{1}\,  v_{a}(\vec{q},\tau_{1}) \, 
B_{b}(\vec{p},\tau_{1}),
\label{NM16}\\
B_{b}^{(1)}(\vec{p},\tau_{1}) &=&  \frac{(-i)}{(2\pi)^{3/2}} \int d^{3} q^{\,\prime} \, \int d^3 p^{\, \prime}  \,\,\, \delta^{(3)}(\vec{p} - \vec{q}^{\, \prime} - \vec{p}^{\, \prime})\,
\nonumber\\
&\times& \epsilon_{b\, m' \, n' \,} \, \epsilon_{a'\, b'\, n'} \, (q_{m ' }' + p_{m' }') \int_{0}^{\tau_{1}} \, d\tau_{2}\,  v_{a'}(\vec{q}^{\,\prime},\tau_{2}) \, 
B_{b}(\vec{p}^{\, \prime}), 
\label{NM17}
\end{eqnarray}
where we used that $ B_{b'}^{(0)}(\vec{p}^{\,\prime},\tau_{2}) = B_{b}(\vec{p}^{\, \prime})$. Inserting Eq. (\ref{NM17}) 
inside Eq. (\ref{NM16}) we have that
 \begin{eqnarray}
 B_{i}^{(2)}(\vec{k},\tau)&=& \frac{(-i)^2}{(2\pi)^{3}}\int d^{3} q \, \int d^3 p\, \int d^{3} q^{\,\prime} \, \int d^3 p^{\, \prime} 
\,\,\, \delta^{(3)}(\vec{k} - \vec{q} - \vec{p})\,\,\, \delta^{(3)}(\vec{p} - \vec{q}^{\, \prime} - \vec{p}^{\, \prime})
 \nonumber\\
 &\times& \int_{0}^{\tau} d\tau_{1} \, \, \int_{0}^{\tau_{1}} d\tau_{2} \, (q_{m} + p_{m}) \, (q_{m ' }' + p_{m' }') 
  \epsilon_{b\, m' \, n' \,} \, \epsilon_{a'\, b'\, n'} \, \epsilon_{m\, n\, i} \, \epsilon_{a\, b\, n}
 \nonumber\\
 &\times& 
  \,  v_{a'}(\vec{q}^{\,\prime},\tau_{2}) \, v_{a}(\vec{q},\tau_{1}) \,B_{b'}(\vec{p}^{\,\prime}).
\label{NM18}
\end{eqnarray}
 By averaging over the velocity fields we have that:
 \begin{eqnarray}
H^{(2)}_{i}(\vec{k},\tau) &=&  \frac{(-i)^2}{(2\pi)^{3}}\int d^{3} q \, \int d^3 p\, \int d^{3} q^{\,\prime} \, \int d^3 p^{\, \prime} 
\,\,\, \delta^{(3)}(\vec{k} - \vec{q} - \vec{p})\,\,\, \delta^{(3)}(\vec{p} - \vec{q}^{\, \prime} - \vec{p}^{\, \prime})
 \nonumber\\
 &\times& \int_{0}^{\tau} d\tau_{1} \, \, \int_{0}^{\tau_{1}} d\tau_{2} \, (q_{m} + p_{m}) \, (q_{m ' }' + p_{m' }') 
  \epsilon_{b\, m' \, n' \,} \, \epsilon_{a'\, b'\, n'} \, \epsilon_{m\, n\, i} \, \epsilon_{a\, b\, n}
 \nonumber\\
 &\times& 
  \,  \langle v_{a'}(\vec{q}^{\,\prime},\tau_{2}) \, v_{a}(\vec{q},\tau_{1}) \rangle \,B_{b'}(\vec{p}^{\,\prime}),
\label{NM19bb}
\end{eqnarray}
where $\langle B_{i}^{(2)}(\vec{k},\tau)\rangle= H^{(2)}_{i}(\vec{k},\tau) $. 
After inserting the explicit expression of the correlator obtained in Eq. (\ref{NM10}), (Eq. (\ref{NM19bb}) reduces to 
 \begin{eqnarray}
H^{(2)}_{i}(\vec{k},\tau) &=&  \frac{(-i)^2}{(2\pi)^{3}}\int d^{3} q \, \int d^3 p\, \int d^{3} q^{\,\prime} \, \int d^3 p^{\, \prime} 
\,\,\, \delta^{(3)}(\vec{k} - \vec{q} - \vec{p})\,\,\, \delta^{(3)}(\vec{p} - \vec{q}^{\, \prime} - \vec{p}^{\, \prime})
 \nonumber\\
 &\times& \int_{0}^{\tau} d\tau_{1} \, \, \int_{0}^{\tau_{1}} d\tau_{2} \, \Gamma(q, \tau_{1},\tau_{2}) \, \frac{q_{a} q_{a'}}{q^2}  \, v(q) \, \delta^{(3)}(\vec{q} + \vec{q}^{\,\prime})
 \nonumber\\
 &\times&\, (q_{m} + p_{m}) \, (q_{m ' }' + p_{m' }') \,  \epsilon_{b\, m' \, n' \,} \, \epsilon_{a'\, b'\, n'} \, \epsilon_{m\, n\, i} \, \epsilon_{a\, b\, n}\,B_{b'}(\vec{p}^{\,\prime}).
\label{NM20}
\end{eqnarray}
Using the three delta functions over the momenta the expression of Eq. (\ref{NM20}) becomes:
 \begin{eqnarray}
H^{(2)}_{i}(\vec{k},\tau) &=&  \frac{(-i)^2}{(2\pi)^{3}}\int d^{3} q \, v(q) B_{b'}(\vec{k}) \, \int_{0}^{\tau} d\tau_{1} \, \, \int_{0}^{\tau_{1}} d\tau_{2} \, \Gamma(q, \tau_{1},\tau_{2}) 
\nonumber\\
&\times& \frac{q_{a} q_{a'}}{q^2} \, k_{m} \, (k_{m'} - q_{m ' }) \, \epsilon_{b\, m' \, n' \,} \, \epsilon_{a'\, b'\, n'} \, \epsilon_{m\, n\, i} \, \epsilon_{a\, b\, n}.
\label{NM21}
\end{eqnarray}
Eq. (\ref{NM21}) can be also written, after some algebra, as
 \begin{eqnarray}
H^{(2)}_{i}(\vec{k},\tau) &=&  \frac{(-i)^2}{(2\pi)^{3}}\int d^{3} q \,\, v(q)\,\, B_{i}(\vec{k}) \,\, \frac{(\vec{k}\cdot\vec{q}) [(\vec{k}\cdot\vec{q}) -q^2]}{q^2} 
\nonumber\\
&\times& \int_{0}^{\tau} d\tau_{1} \, \, \int_{0}^{\tau_{1}} d\tau_{2} \, \Gamma(q, \tau_{1},\tau_{2}).
\label{NM22}
\end{eqnarray}
By appreciating that $ \Gamma(q,\tau_{1},\tau_{2}) = \Gamma(q,\tau_{2},\tau_{1})$  the integrations over $\tau_{1}$ and $\tau_{2}$ can be explicitly performed by recalling the elementary integral 
\begin{equation}
 \int_{0}^{y} \sin{(b x)}\,\, e^{- a x} \,\, d x= \frac{b}{a^2 + b^2} - \frac{b}{a^2 + b^2} \biggl[ 
 \cos{b y} + \frac{a}{b} \sin{b y} \biggr] e^{- a y}.
 \label{NM25}
 \end{equation}
The result is:
\begin{equation}
\overline{\gamma}(q,\tau) =\int_{0}^{\tau} d\tau_{1} \, \, \int_{0}^{\tau_{1}} d\tau_{2} \, \Gamma(q, \tau_{1},\tau_{2})  = \frac{\{ 1 - e^{- q^2 \nu_{\mathrm{th}} \tau} [\cos{(q c_{\mathrm{sb}} \tau)} + \epsilon(q,\tau) \sin{(q c_{\mathrm{sb}} \tau)} ]\}^2}{2 q^2 
 c_{\mathrm{sb}}^2 [ 1 + \epsilon^2(q,\tau)]^2},
 \nonumber
 \end{equation}
 where $\epsilon(q,\tau) =  q^2 \nu_{\mathrm{th}}^2 / c_{\mathrm{sb}}^2  \ll 1$.  In the 
 limit $\epsilon(q,\tau) \ll 1$ and $q^2 \nu_{\mathrm{th}} \tau$
 $\gamma(q\tau)\simeq [c_{\mathrm{sb}}^2 \tau^4]/8$ and the higher order contributions 
 \begin{eqnarray}
 H_{i}^{(2)}(\vec{k},\tau) &=& - \frac{k^2\, B_{i}(\vec{k})}{3} \,\int \frac{d^3 q}{(2\pi)^3} \, \gamma(q, \tau) v(q),\qquad 
 \nonumber\\
 H_{i}^{(4)}(\vec{k},\tau) &=& \frac{1}{2!} \frac{k^2\, B_{i}(\vec{k})}{3} \,\int \frac{d^3 q}{(2\pi)^3} \, \gamma(q, \tau) v(q)\,\int \frac{d^3 p}{(2\pi)^3} \, \gamma(p, \tau) v(p),
 \nonumber\\
  H_{i}^{(6)}(\vec{k},\tau) &=& -\frac{1}{3!} \frac{k^2\, B_{i}(\vec{k})}{3} \,\int \frac{d^3 q}{(2\pi)^3} \, \gamma(q, \tau) v(q)\,\int \frac{d^3 p}{(2\pi)^3} \, \gamma(p, \tau) v(p)
 \nonumber\\ 
&\times&  \,\int \frac{d^3 k}{(2\pi)^3} \, \gamma(k, \tau) v(k),
 \label{NM26}
 \end{eqnarray}
and so on can be resummed leading to an average magnetic field
 \begin{equation}
H_{i}(\vec{k},\tau) = e^{- f (k,\tau)}\, B_{i}(\vec{k}), \qquad f(k,\tau) = 
\frac{c_{\mathrm{sb}}^2}{24 (n + 1) } {\mathcal A}_{{\mathcal R}}\, k_{\mathrm{p}}^2 \, k^2 \tau^4 \, \biggl(\frac{k_{\mathrm{d}}}{k_{\mathrm{p}}}\biggr)^{n + 1}.
\label{NM27}
\end{equation}
 The initial magnetic field is suppressed when $f(k,\tau_{*}) \geq 1$, i.e. when
 \begin{equation}
 \frac{k_{\mathrm{nM}}}{k_{\mathrm{p}}} \geq 20 \frac{\sqrt{2 (n + 1)}}{c_{\mathrm{sb}}(d_{\mathrm{A}} k_{\mathrm{p}})^2} {\mathcal A}_{{\mathcal R}}^{-1/2}  \biggl(\frac{k_{\mathrm{p}}}{k_{\mathrm{d}}}\biggr)^{(n +1)/2},
 \label{NM28}
 \end{equation} 
where $d_{\mathrm{A}}$ is the (comoving) angular diameter distance to last scattering and $k_{\mathrm{nM}}$ denotes the diffusion scale in the non-Markovian approach proposed in this appendix.
The results obtained here are consistent 
with a rough dimensional going, in short, as follows. Equation (\ref{NM11}) does not possess stationary solutions (see, e.g. \cite{vain0}): the field is either amplified or dissipated. The typical diffusivity scale induced by large-scale curvature perturbations can be simply obtained 
by balancing the left and the right-hand sides of Eq. (\ref{NM11}) and by assuming that the typical amplitude of the velocity 
field is given by $v_{\gamma\mathrm{b}} \sim \sqrt{{\mathcal A}_{{\mathcal R}}}$ (for $n_{\mathrm{s}} \simeq 1$). In this case the typical 
diffusivity length is $L \sim \sqrt{{\mathcal A}_{{\mathcal R}}} \tau \sim  \sqrt{{\mathcal A}_{{\mathcal R}}} {\mathcal H}^{-1}$.
\end{appendix}

\end{document}